\def\be{\begin{equation}}
\def\ee{\end{equation}}
\def\bea{\begin{eqnarray}}
\def\eea{\end{eqnarray}}
\def\nn{\nonumber}
\def\qtwo{\qquad\qquad}

\def\eqdef{\stackrel{\rm def}{=}}
\def\bigoh{{\mathcal O}}

\def\Cov{\mbox{Cov}}
\def\Var{\mbox{Var}}
\def\Corr{\mbox{Corr}}
\def\tC{{\widetilde C}}
\def\tT{{\widetilde T}}
\def\tE{{\widetilde E}}
\def\bl{{\bf l}}
\def\G{{\mathcal G}}
\def\N{{\mathcal N}}
\def\hbn{\widehat{\bf n}}
\def\bx{{\bf x}}
\def\tQ{{\widetilde Q}}
\def\tU{{\widetilde U}}
\def\tE{{\widetilde E}}
\def\tB{{\widetilde B}}
\def\hD{{\widehat\Delta}}

\def\hE{{\widehat{\mathcal E}}}
\def\hDelta{{\widehat{\Delta}}}
\def\hpi{{\widehat\pi}}

\def\ba{{\rm lo}}
\def\bb{{\rm hi}}
\def\wm{\Omega_c h^2}
\def\wn{\sum m_\nu}
\def\ldz{\ln\,\delta_\zeta}
\def\ellmax{l_{\rm max}}
\def\fsky{f_{\rm sky}}

\newcommand{\tot}{{\rm eff}}
\newcommand{\meas}{}

\documentclass[twocolumn,amsmath,amssymb,prd]{revtex4}
\usepackage{bm}
\usepackage{epsf}

\begin{document}

\title{Cosmological Information from Lensed CMB Power Spectra}
\author{Kendrick M. Smith$^{1,2}$, Wayne Hu$^{1,3}$ and Manoj Kaplinghat$^{4}$} 
\affiliation{
$^{1}$Kavli Institute for Cosmological Physics, Enrico Fermi Institute, University of Chicago, 60637\\
$^{2}$Department of Physics,  University of Chicago, 60637\\
$^{3}$Department of Astronomy and Astrophysics, University of Chicago, 60637\\
$^{4}$Center for Cosmology, Dept, of Physics \& Astronomy, University of California, Irvine CA}

\begin{abstract}
Gravitational lensing distorts the cosmic microwave background (CMB)  temperature and polarization fields and encodes valuable
information on distances and growth rates at intermediate redshifts into the lensed power spectra.
The non-Gaussian bandpower covariance induced by the lenses
 is negligible to $l=2000$ for all but the $B$ polarization field
where it increases the net variance by
up to a factor of 10 and favors an observing strategy with 3 times more area than if it were Gaussian.
To quantify the cosmological information, we introduce two lensing observables, characterizing nearly
all of the information, which
simplify the study of non-Gaussian impact, parameter degeneracies, dark energy models,
and complementarity with other cosmological probes.   
Information on the  intermediate redshift parameters 
rapidly becomes limited by constraints on the cold
dark matter density and initial amplitude of fluctuations as observations improve.   
Extraction of this information requires deep polarization
measurements on only 5-10\% of the sky, and can improve Planck lensing constraints by
a factor of $\sim 2-3$ on any {\em one} of the parameters \{$w_0,w_a,\Omega_K,\sum m_\nu$\} 
with the others fixed.   
Sensitivity to the curvature and neutrino mass are the highest due to the high redshift
weight of CMB lensing but degeneracies between the parameters must be broken externally.
\end{abstract}

\maketitle

\section{Introduction}

Primary cosmic microwave background (CMB) anisotropy from recombination has 
proven itself to be a veritable gold mine of cosmological information.
One of the most important secondary signals that should be detected by upcoming cosmic microwave
background experiments
is the distortion to the temperature and polarization fields due to
 gravitational lensing by the large-scale
structure of the universe (see \cite{ChaLew05} for a recent review).  Lensing
distortions
add cosmological information  on parameters such as curvature, neutrino masses
and dark energy that change the expansion and growth rate at
intermediate redshifts ($z \lesssim 5$). 
 
 This distortion in real space couples power in harmonic space
 and hence introduces non-Gaussianity into the CMB temperature
 and polarization fields.  Beyond power spectra, this
 non-Gaussianity is a source of information in that it 
 allows direct reconstruction of the convergence
 field \cite{Ber98,ZalSel99,Hu01b,HuOka01,HirSel02}.
On the other hand, for purposes of extracting cosmological
 information from lensed power spectra as considered here, this non-Gaussianity is
 largely an impediment as it makes power spectrum estimates
 covary across a wide range of multipoles.

The purpose of this paper is twofold.  First, we calculate the full
non-Gaussian covariance between all combinations of temperature and
polarization bandpowers in the lensed CMB.
This extends previous work in which the
temperature \cite{Hu01,Coo02} and $B$-mode polarization covariance \cite{SmiHuKap04}
were calculated separately.
Second, we present a general framework for studying the extra information on
cosmological parameters that lensed CMB spectra supply,
with particular attention to the impact of
non-Gaussianity.    

Previous works have noted that the lensed CMB signal may be used to
study the dark energy \cite{Hu01c,Kap03,AcqBac05} and neutrino mass
\cite{KapKnoSon03}. These studies did not compute the non-Gaussian
covariance but assumed either that the information is encoded in the
unlensed primary CMB and a reconstruction of the lenses or by approximating
the non-Gaussian covariances with a degradation factor from \cite{SmiHuKap04}. 
Our results lend support to these analyses. 
We also study the sensitivity of lensing to curvature and find that 
future CMB measurements can provide interesting constraints on it.  

This paper is organized as follows.
In \S\ref{sec:pscov}, we compute non-Gaussian contributions to the covariance
between all lensed CMB temperature and polarization bandpowers. We then describe
in \S\ref{sec:formalism} how this non-Gaussian covariance propagates 
into Fisher matrix parameter forecasts and present formal bounds on its impact.
In \S\ref{sec:deobs}, we define two parameter independent observables which contain
essentially all information from the lensed CMB and discuss their relationship to distance and
 growth as well as their degeneracy with parameters
that control the matter power spectrum.
Armed with this general framework, we show how constraints on these observables can be 
interpreted in the context of common parameterizations of the dark energy and dark matter in
\S\ref{sec:exfisher}.
Finally in \S\ref{sec:applications} we show how 
 future CMB surveys can be optimized for sensitivity to the lensing observables.
We conclude in \S\ref{sec:discussion} and briefly address the issues of 
 goodness-of-fit in Appendix \ref{sec:chisq} and scaling with the fiducial cosmology in
 Appendix \ref{sec:fiducial}.

\section{Lensed power spectrum covariance}
\label{sec:pscov}

In this section, we compute the non-Gaussian covariance between all
CMB temperature and polarization bandpowers to lowest
order in the lensing power spectrum $C_l^{\phi\phi}$.
The results (Eqs.~(\ref{eq:bbbb}), (\ref{eq:xybb}), and (\ref{eq:xyzw}))
will be foundational in subsequent sections, as they will permit the 
effects of non-Gaussianity to incorporated into parameter forecasts.
However, the details of the calculation will not be needed,
so the reader may wish to skip this section on a first reading.

First, we recall some preliminaries concerning lensed CMB fields.
We work in the flat sky approximation; we 
will see (\S\ref{sec:psest}) that non-Gaussian covariance only becomes important when
combining bandpowers over a wide range of $l$, 
so that all-sky corrections from the discrete nature of $l$ should be negligible.
The lensed CMB temperature $T(\bx)$ and unlensed temperature
$\tT(\bx)$ are related by 
\be
T(\bx) = \tT(\bx + \nabla\phi(\bx)) \,.
\label{eq:Tdef}
\ee
The projected potential $\phi$ is given by the line-of-sight integral:
\be
\phi(\hbn) = 2 \int dD \frac{D_A(D_s - D)}{D_A(D) D_A(D_s)} \Phi(D\hbn,D)\,,
\label{eq:projectedpotential}
\ee
where $D=\int dz/H$ is the comoving distance along the line of sight,
$D_s$ denotes the comoving distance to the surface of last scattering, and
\be
D_A(D) = {1\over \sqrt{\Omega_K H_0^2}} \sinh \left( \sqrt{\Omega_K H_0^2} D\right) 
\ee
is the comoving angular diameter distance.

Polarization fields are lensed in the same way; the lensed Stokes parameters $Q(\bx)$, $U(\bx)$
and unlensed versions $\tQ(\bx)$, $\tU(\bx)$ are related by
\bea
Q(\bx) &=& \tQ(\bx + \nabla\phi(\bx))\,,  \nn \\
U(\bx) &=& \tU(\bx + \nabla\phi(\bx))\,.  \label{eq:Udef}
\label{eql1}
\eea
The Fourier versions of Eqs. (\ref{eq:Tdef}), (\ref{eq:Udef}) are \cite{Hu00b}:
\bea
T(\bl) &=& \tT(\bl) + \int \frac{d^2\bl'}{(2\pi)^2} W_T(\bl,\bl') \tT(\bl') \phi(\bl-\bl') + \bigoh(\phi^2)\,, \nn  \\
E(\bl) &=& \tE(\bl) + \int \frac{d^2\bl'}{(2\pi)^2} W_E(\bl,\bl') \tE(\bl') \phi(\bl-\bl') + \bigoh(\phi^2)\,, \nn \\
B(\bl) &=& \int \frac{d^2\bl'}{(2\pi)^2} W_B(\bl,\bl') \tE(\bl') \phi(\bl-\bl') + \bigoh(\phi^2)\,, \label{eq:fouriervers}
\eea
where the kernels are defined by
\bea
W_T(\bl,\bl') &=& -[\bl'\cdot (\bl-\bl')]   \,,  \nn \\
W_E(\bl,\bl') &=& -[\bl'\cdot (\bl-\bl')] \cos 2(\varphi_{\bl}-\varphi_{\bl'})\,,  \nn \\
W_B(\bl,\bl') &=&  [\bl'\cdot (\bl-\bl')] \sin 2(\varphi_{\bl}-\varphi_{\bl'})\,, \label{eq:kerndef}
\eea
and the $E$ and $B$ fields are related to the Stokes parameters as
\be
[E(\bl) \pm iB(\bl)] = [Q(\bl) \pm iU(\bl)] \exp(\mp 2i\varphi_{\bl}) \,.
\ee
Here $\varphi_{\bl}$ is the angle between ${\bf l}$ and $\hat{\bf x}$ axis.  We
have assumed that the unlensed $\tB=0$ such that the observed $B$
field is generated from $\tE$ by lensing alone \cite{ZalSel98}.

We define ideal, noise-free bandpower estimators by
\be
\hD_i^{XY} = \frac{1}{A\alpha_i} \int_{\bl\in i} d^2\bl\, \left(\frac{l^2}{2\pi}\right) X^*(\bl) Y(\bl)\,,
\label{eq:Ddef}
\ee
where $XY \in \{TT,EE,TE,BB\}$, $A$ is the survey area in steradians, and
\be
\alpha_i = \int_{\bl\in i} d^2\bl
\ee
is the $l$-space area of band $i$. We define power spectra as usual
\bea
\langle X^*(\bl) Y(\bl') \rangle &=& (2\pi)^{2} \delta^2({\bf l}-{\bf l'}) C_{\bl}^{XY} \nn\\
&\approx&  A {\delta_{{\bf l},{\bf l'}}} C_{\bl}^{XY} \,,
\eea
such that 
\be
\Delta_{i}^{XY} \eqdef \langle \hD_{i}^{XY} \rangle = 
\frac{1}{\alpha_{i}} \int_{\bl\in i} d^2\bl\,  \left(\frac{l^2}{2\pi}\right)  C_{\bl}^{XY} \,.
\ee
(As a technical point, when propagating $C^{\phi\phi}_l$ to lensed power spectra, we use the
all-sky correlation function approach of \cite{ChaLew05}, for consistency with CAMB.)

Now let us consider the covariance of these estimators in non-overlapping $l$ bands.
We split the bandpower covariance into Gaussian and non-Gaussian pieces:
\bea
C^{IJ} &\eqdef &  %&\Cov(\hD_i^{XY},\hD_j^{ZW}) \nn\\ 
\langle \hD_{i}^{XY}\hD_{j}^{ZW} \rangle  - \Delta_{i}^{XY}\Delta_{j}^{ZW} \nn\\
&=& \G_{ij}^{XY,ZW} + \N_{ij}^{XY,ZW} \,.
\label{eq:GNdef}
\eea
Here and below we will use the short hand notation $I$ to denote a unique bandpower
specified by the $l$-band $i$ and the power spectrum $XY$.
The Gaussian piece is given by:
\bea
\G_{ij}^{XY,ZW} &=& \delta_{ij} \frac{(2\pi)^2}{A\alpha_i^2} \int_{\bl\in i} d^2\bl\, \left( \frac{l^2}{2\pi} \right)^2  \\
                 && \qtwo \times ( C_{\bl}^{XZ}C_{\bl}^{YW} + C_{\bl}^{XW}C_{\bl}^{YZ} )\,,  \nn
\label{eq:Gdef}
\eea
where the power spectra which appear are lensed.  In the presence of 
 instrumental noise, the power spectra in this formula are replaced as
 \bea
 C_{\bl}^{XY} &\rightarrow&  C_{\bl}^{XY} + N_{\bl}^{XY} \,,
 \eea
 where the noise power spectra for white detector noise with a Gaussian beam are given by
 \bea
 N_{\bl}^{XX} = \left(  {\Delta_X \over T_{\rm CMB}} \right)^2 e^{l(l+1)\theta_{\rm FWHM}^2/8\ln 2} \,,
 \label{eq:NLdef}
 \eea
 for $XX \in TT, EE, BB$ and vanishing for other spectra.  We will also take $\Delta_E=\Delta_B
 =\Delta_P$.

In \citep{SmiHuKap04}, we computed the non-Gaussian piece $\N_{ij}^{XY,ZW}$ for the case of two $BB$ bandpowers:
\bea
\N_{ij}^{BB,BB} &=& \frac{2}{A\alpha_i\alpha_j} \int_{\bl_i\in i} d^2\bl_i \int_{\bl_j\in j} 
                    d^2\bl_j \int \frac{d^2\bl}{(2\pi)^2} \label{eq:bbbb} \\
                 && \qquad \times \frac{l_i^2 l_j^2}{(2\pi)^2}
                  (a_{\bl_i\bl_j}^{\bl} + b_{\bl_i\bl_j}^{\bl} + c_{\bl_i\bl_j}^{\bl}) + 
                  \bigoh(C_l^{\phi\phi})^3\,, \nn
\eea
where
\bea
a^{\bl}_{\bl_i\bl_j} &=& W^2_B(\bl_i,\bl_i-\bl) W_B^2(\bl_j,\bl_j-\bl) \tC^{EE}_{\bl_i-\bl}
                         \tC^{EE}_{\bl_j-\bl} \big( C^{\phi\phi}_{\bl} \big)^2\,,   \nn \\
b^{\bl}_{\bl_i\bl_j} &=& W^2_B(\bl_i,\bl) W^2_B(\bl_j,\bl) \big( \tC_{\bl}^{EE} \big)^2
                         C^{\phi\phi}_{\bl_i-\bl} C^{\phi\phi}_{\bl_j-\bl}  \,,       \\
c^{\bl}_{\bl_i\bl_j} &=& W_B(\bl_i,\bl_i-\bl) W_B(-\bl_i,\bl_j-\bl) W_B(\bl_j,\bl_j-\bl) \nn \\
                      && \times W_B(-\bl_j,\bl_i-\bl) \tC^{EE}_{\bl_i-\bl} 
                         \tC^{EE}_{\bl_j-\bl} C^{\phi\phi}_{\bl} C^{\phi\phi}_{\bl_i+\bl_j-\bl}\,, \nn
\eea
with $\tC^{XY}_l \eqdef C^{\tilde X \tilde Y}_l$ as the unlensed power spectra.

Here, we consider two additional cases.
First, we compute the non-Gaussian covariance of one $BB$ bandpower with a bandpower $\hD_i^{XY}$, where $X,Y\in\{T,E\}$:
\bea
\N_{ij}^{XY,BB} &=& \frac{2}{A\alpha_i\alpha_j} \int_{\bl_i\in i} d^2\bl_i \int_{\bl_j\in j} 
                    d^2\bl_j \frac{l_i^2 l_j^2}{(2\pi)^2} \,,\nn \\
                 && \qquad \times  W_B(\bl_j,\bl_i)^2 \tC^{EX}_{\bl_i} \tC^{EY}_{\bl_i} 
                    C^{\phi\phi}_{\bl_i-\bl_j} \,. \label{eq:xybb}
\eea
Second, we compute the covariance of two bandpowers $\hD_i^{XY}$, $\hD_j^{ZW}$, where
$X,Y,Z,W \in \{T,E\}$:
\bea
\N_{ij}^{XY,ZW} &=& \frac{1}{A\alpha_i\alpha_j} \int_{\bl_i\in i} d^2\bl_i \int_{\bl_j\in j} d^2\bl_j 
                  \frac{l_i^2 l_j^2}{(2\pi)^2} C_{\bl_i-\bl_j}^{\phi\phi}  \label{eq:xyzw} \\
                && \qquad \times (\alpha_{\bl_i\bl_j}^{XY,ZW} + \alpha_{\bl_j\bl_i}^{ZW,XY} 
                   + \beta_{\bl_i\bl_j}^{XY,ZW}  \nn \\
                && \qtwo + \beta_{\bl_i\bl_j}^{XY,WZ} + \beta_{\bl_i\bl_j}^{YX,ZW} + \beta_{\bl_i\bl_j}^{YX,WZ}). \nn
\eea
Here,
\bea
\alpha_{\bl_i\bl_j}^{XY,ZW} \!\!\!\!&=& W_X(\bl_i,\bl_j) W_Y(\bl_i,\bl_j) 
                                (\tC_{\bl_j}^{XZ}\tC_{\bl_j}^{YW} \!\! + \tC_{\bl_j}^{XW}\tC_{\bl_j}^{YZ})\,,  \nn \\
\beta_{\bl_i\bl_j}^{XY,ZW} \!\!\!\! &=& W_X(\bl_i,\bl_j) W_Z(\bl_j,\bl_i) \tC^{XW}_{\bl_j} \tC^{YZ}_{\bl_i}\,.
\eea
Taken together, Eqs. (\ref{eq:bbbb}), (\ref{eq:xybb}), (\ref{eq:xyzw}) 
constitute a complete calculation of the $(4N_{\rm band})$-by-$(4N_{\rm band})$
covariance matrix.

In Table \ref{tab:mc}, we show the correlations
\be
R^{IJ} = \frac{C^{IJ}}{\sqrt{ C^{II} C^{JJ}}}
\ee
and variance degradation factors
\be
D^{I} = \frac{({\bf C})^{II}}{({\bf C}^{G})^{II}}
\label{eq:vardegradation}
\ee
for all combinations of \{$TT, TE, EE, BB$\}
in two large $l$ bands, with Monte Carlo results from $10^5$ simulations shown for comparison.  
Here ${\bf C}^{\rm G}$ represents the Gaussian contribution to the covariance matrix.
Throughout this paper, we use a fiducial model consistent with the third-year
WMAP \cite{Speetal06} data:
\bea
 \{ \Omega_b h^2, \Omega_c h^2, \tau, \ldz, n_s, r \} &=&
  \{ 0.0223, 0.104,  0.088, \nn\\
  && \quad  -10, 0.951, 0 \} \,, \label{eq:fidmodel}  \\
  \{ \Omega_{\rm DE}, \Omega_{K}, \wn, w_{0}, w_{a} \} &=&
 \{ 0.76, 0, 0.061{\rm eV}, -1, 0 \}\,. \nn
\eea
The first set represents parameter that control the intrinsic power spectra from recombination
whereas the second set represents the intermediate redshift parameters of interest to
lensing.
Here $\delta_\zeta \propto \sqrt{A}$ is the amplitude of initial curvature fluctuations at $k=0.05$ Mpc$^{-1}$ and $r$ is the tensor to scalar ratio.  The dark energy (DE) equation of state
is parameterized as 
\be
w(a) = w_0 + (1-a) w_a \,.
\label{eq:wa}
\ee
When $w_a=0$ we will use the variables $w$ and $w_0$ interchangeably.

The most important part of the non-Gaussian covariance is between two
$BB$ bandpowers, where the analytic and Monte Carlo results agree well.  The second most
important contribution is the correlations between $EE$ and $BB$ which appear at the 10-30\% level.  
Here the agreement between
Eq.~(\ref{eq:xybb}) and the Monte Carlo results for the correlation are at the 
10-20\% level indicating that
higher order contributions are  not entirely negligible.  However they represent a small correction to
a small correlation and we neglect it throughout.  In the remaining parts 
of the covariance matrix, including the entire \{$TT, TE, EE$\} covariance, the non-Gaussian 
contributions are small.

Since the non-Gaussianity
manifests itself as a small correlation across a wide range of multipoles, it is only
visually apparent when combining the multipoles into large bands as in Tab. \ref{tab:mc}.
In practice, in the following sections we take bins of $\Delta l= 10$ out to $l_{\rm max}=2000$,
and all results in the paper are robust to binning more finely.

\begin{table*}
\begin{center}
\begin{tabular}{|c|cccccccc||c|}
\hline    & $TT_\ba$ & $TT_\bb$ & $TE_\ba$ & $TE_\bb$ & $EE_\ba$ & $EE_\bb$ & $BB_\ba$ & $BB_\bb$ &  $D$    \\  \hline
$TT_\ba$  &     1    &  0.007   &  -0.053  &  0.001   &  0.074   &  0.001   &  0.025   &  0.009   &  1.007 (1.012)  \\
$TT_\bb$  &  (0.008) &     1    &  0.001   &  -0.312  &  0.003   &  0.089   &  0.014   &  0.025   &  1.020 (1.019) \\
$TE_\ba$  &  (-0.055)&  (0.002) &    1     &  0.003   &  -0.098  &  0.001   &  -0.036  & -0.010   &  1.000 (1.000) \\
$TE_\bb$  &  (0.001) &  (-0.311)& (0.004)  &    1     &  0.001   &  -0.306  &  -0.049  & -0.086   &  1.010 (1.011) \\
$EE_\ba$  &  (0.076) &  (-0.001)& (-0.096) & (0.001)  &    1     &  0.004   &  0.316   &  0.137   &  1.012 (1.011) \\
$EE_\bb$  &  (0.002) &  (0.090) & (0.002)  & (-0.311) & (0.003)  &    1     &  0.137   &  0.283   &  1.039 (1.039) \\
$BB_\ba$  &  (0.022) &  (0.027) & (-0.048) & (-0.030) & (0.311)  & (0.117)  &     1    &  0.754   &  4.323 (4.416) \\
$BB_\bb$  &  (0.005) &  (0.039) & (-0.021) & (-0.067) & (0.132)  & (0.262)  &  (0.754) &    1     &  7.595 (7.619) \\  \hline
\end{tabular}
\end{center}
\caption{\footnotesize Bandpower correlations and variance degradation factors
$D$ (see Eq.~(\ref{eq:vardegradation})) using two bands in $l$: (lo)$=[100,1000]$ and (hi)$=[1000,2000]$.
Upper diagonal values were calculated analytically using the lowest-order expressions
Eqs. (\ref{eq:bbbb}), (\ref{eq:xybb}), (\ref{eq:xyzw}); lower diagonal values (in parentheses)
were calculated from $10^5$ Monte Carlo
simulations.  The large correlations between TE and \{$TT$, $EE$\} are dominated by the Gaussian
contribution.}

\label{tab:mc}
\end{table*}

\section{Parameter Forecast Formalism}
\label{sec:formalism}

In this section, we review the Fisher matrix formalism for forecasting parameter 
errors in the presence of non-Gaussian errors in \S \ref{sec:fisher}.   We then apply  this formalism to place 
formal bounds on the impact of non-Gaussianity in \S \ref{sec:psest}.

\subsection{Fisher matrix}
\label{sec:fisher}

The Fisher matrix provides a useful way of assessing the impact of the non-Gaussian
bandpower 
covariance on parameter estimation.  Even for cases where the likelihood function 
cannot be evaluated directly,
it can be approximated as the linear propagation of
errors from bandpower space to another parameter space $p_\alpha$.   
 In \S\ref{sec:pscov}, we gave a complete calculation, to lowest order in
$C_l^{\phi\phi}$, of the $(4N_{\rm band})$-by-$(4N_{\rm band})$ covariance ${\bf C}^{IJ}$ between bands 
$I$ and $J$ specified by the power spectrum combination and the $l$ range (see Eq.~(\ref{eq:GNdef})).
In terms of this covariance, 
%the Fisher matrix is given by
we define an approximate Fisher matrix as
\be
F_{\alpha\beta} =\sum_{IJ} (\partial_\alpha \Delta^I) ({\bf C}^{-1})_{IJ} (\partial_\beta \Delta^J)\,,
\label{eq:Fdef}
\ee
where $\alpha$, $\beta$ run over a basis set of directions in parameter space.
 In this section we
will use upper indices to denote quantities that transform as a contravariant tensor under
a re-parameterization and lower indices for those that transform as a covariant tensor.
For example, Eq.~(\ref{eq:Fdef}) represents the transformation of the inverse covariance
matrix from the bandpower space to the parameter basis space.

Given this transformation, the inverse of the Fisher matrix 
can be thought of as an estimate of 
the covariance matrix of the basis parameters.
As such it gives the variance of any linear combination of 
basis parameters as 
\be
\Var(\pi) = \sum_{\alpha\beta}(\partial_\alpha \pi) ({\bf F}^{-1})^{\alpha\beta} (\partial_\beta \pi)\,.
\label{eq:perr}
\ee
As a special case, if $\pi$ corresponds to a basis direction $\alpha$, then the marginalized uncertainty is given by
the diagonal element $({\bf F}^{-1})^{\alpha\alpha}$.

The Fisher matrix quantifies the local curvature of
the likelihood function in the parameter space.  
 Fisher forecasts therefore suffer from several problems (see {\em e.g.},
\cite{EisHuTeg99b,PerLesHanTuWon06}), especially in the presence of nearly
degenerate  parameter directions.   If the derivatives in
Eq.~(\ref{eq:Fdef}) are not 
constant across the extent of the degeneracy, this curvature is also not constant and hence
confidence regions in parameter space are not well approximated by ellipsoids in either their shape
or extent.     Fisher matrix forecasts
should only be interpreted as confidence limits on parameters for well-constrained
directions in the parameter space and as a tool to expose parameter degeneracies.
Finally even if the parameter derivatives vary significantly only outside of the predicted error 
ellipsoid, Fisher matrix forecasts still depend on the choice of the fiducial model. 
These points should be kept in mind when considering the parameter forecasts in the next
two sections.

\subsection{Formal Bounds on Non-Gaussian Impact}
\label{sec:psest}

Before turning to parameter forecasts in specific parameter spaces,  it is instructive to
quantify general bounds on the impact of non-Gaussianity.
One of the main results of this paper is that non-Gaussian power spectrum covariance is essentially
negligible when considering lensed \{$T,E$\} alone at $\ellmax = 2000$ (c.f. \cite{Hu01} for higher $\ellmax$).
Beyond this $\ellmax$ other secondary sources of temperature and polarization will likely
prohibit the extraction of cosmological information.

To state this in a precise way, we introduce Karhunen-Lo\`eve (KL) eigenvalues between the non-Gaussian and
Gaussian bandpower covariances.  The KL eigenvalues $\lambda_K$ and
eigenvectors $v^K$ are defined by  
\be
\sum_J ({\bf C})^{IJ} (v^K)_J = \lambda_K \sum_J ({\bf C}^{\rm G})^{IJ} (v^K)_J \,,
\label{eq:lambdadef}
\ee
where ${\bf C}$, ${\bf C}^{\rm G}$ denote the bandpower covariances with and without non-Gaussian contributions.
With cosmic variance limited \{$TT,TE,EE$\} bandpowers to $l_{\rm max}=2000$, we
find that each $\lambda_K$ is between $\lambda_{\rm min}$ = 0.94 and $\lambda_{\rm max}$ = 1.08
in the fiducial cosmology.  The exact values depend on the normalization of the power spectrum
but remain around unity for all reasonable variations.

These eigenvalues limit the excess variance on parameter errors from non-Gaussianity.
More precisely, we now prove that for any cosmological parameter $\pi$, the ratio of
uncertainties with and without including non-Gaussian covariance satisfies:
\be
\lambda_{\rm min} \le \frac{\Var(\pi)}{\Var^{\rm G}(\pi)} \le \lambda_{\rm max}\,. \label{eq:keyineq}
\ee
This inequality holds even after marginalizing any set of additional parameters.

The first step in the proof of Eq.~(\ref{eq:keyineq}) is to note that in a basis consisting of
KL eigenvectors in the bandpowers,
both the non-Gaussian and Gaussian bandpower covariance matrices are diagonal:
\be
({\bf C})^{KK'} = \lambda_K \delta_{K K'}\,  \qquad   ({\bf  C}^{\rm G})^{KK'}= \delta_{K K'}\,.   \label{eq:KLproof1}
\ee
(More generally, this holds true for symmetric matrices  as long as one of them is
positive-definite.)
Now consider any estimator $\hE$ which is linear in the bandpowers.  In the KL basis, it can be written
\be
\hE = \sum_K (\hE_K) \hDelta^K   \,. \label{eq:KLproof2}
\ee
Combining Eqs.~(\ref{eq:KLproof1}) and (\ref{eq:KLproof2}),
the ratio between the estimator variance calculated with and without non-Gaussian covariance satisfies:
\be
\frac{\Var(\hE)}{\Var^{\rm G}(\hE)} = \frac{\sum_K \lambda_K (\hE_K)^2}{\sum_K (\hE_K)^2}\,.
\ee
In this form, it is seen that
\be
\lambda_{\rm min} \le \frac{\Var(\hE)}{\Var^{\rm G}(\hE)} \le \lambda_{\rm max}\,.   \label{eq:KLproof3}
\ee
Next we observe that in the Fisher approximation, the estimator for any cosmological parameter $\pi$
depends linearly on the estimated bandpowers.
(This is still true if additional parameters are marginalized, although marginalization will change the optimal
estimator.) 
We denote the estimator which is optimal with the non-Gaussian covariance included
by $\hpi$ and that without $\hpi_{\rm G}$.  Then:
\be
\lambda_{\rm min} \le
\frac{\Var(\hpi)}{\Var^{\rm G}(\hpi)} \le
\frac{\Var(\hpi)}{\Var^{\rm G}(\hpi_{\rm G})} \le
\frac{\Var(\hpi_{\rm G})}{\Var^{\rm G}(\hpi_{\rm G})} \le
\lambda_{\rm max}\,,
\ee
where we have combined Eq.~(\ref{eq:KLproof3}) with the inequalities
$\Var^{\rm G}(\hpi_{\rm G})\le\Var^{\rm G}(\hpi)$ and $\Var(\hpi)\le\Var(\hpi_{\rm G})$, which follow from the
optimality of each estimator.
This completes the proof of~Eq.~(\ref{eq:keyineq}).

The KL eigenvectors also illuminate the nature of the non-Gaussian covariance.
The first (largest $\lambda$) eigenvector is the combination
of bandpowers whose variance degrades the most (relative to Gaussian) when non-Gaussianity is included;
the second eigenvector degrades the second most, and so on.
For \{$TT,TE,EE$\}, all eigenvalues are close to 1, and the variance degradation is essentially negligible,
as quantified by the inequality in Eq.~(\ref{eq:keyineq}), in any direction in parameter space.

\begin{figure}
\centerline{\epsfxsize=3.0truein\epsffile[80 520 320 700]{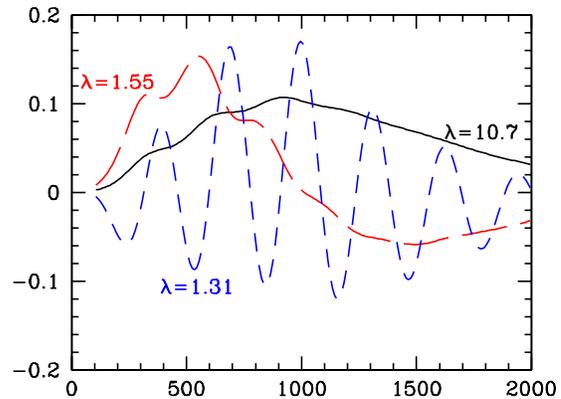}}
\caption{\footnotesize First three KL eigenmodes, defined by Eq.~(\ref{eq:lambdadef}), for the $BB$ power spectrum.
These represent modes in $BB$ whose true variance is larger than the variance estimated from Gaussian statistics;
the eigenvalue $\lambda$ is the ratio of the two.}
\label{fig:eigbb}
\end{figure}

For $BB$ bandpowers, we show the first few eigenvalues and eigenvectors in Figure \ref{fig:eigbb}.
The main effect of the non-Gaussian covariance is to degrade the variance, by a factor of $\sim 10$,
for one KL component which is coherent across a wide range of $l$ and has roughly
the same shape as the fiducial $BB$ spectrum.
This is consistent with \citep{SmiHuKap04}, in which we found that non-Gaussianity degraded the uncertainty
in the overall amplitude of the $BB$ spectrum by a factor of $\sim 10$ when sample variance limited to
$l_{\rm max}=2000$.
Here we see that this single statement roughly characterizes the entire non-Gaussian covariance between
\{$TT, TE, EE, BB$\} bandpowers.

\section{CMB Lensing Observables}
\label{sec:deobs}

In this section, we quantify the information in terms of the power spectrum of the lenses.
An examination of the
information contained in the power spectrum of the lenses serves a dual purpose: 
 it is a parameterization independent quantification of
the additional information from lensing (\S \ref{sec:pcomp}, \ref{sec:redshift}), and
it exposes the origin of the non-Gaussian covariance of the CMB as arising from the sample variance
of the lenses (\S \ref{sec:lensvariance}).

\begin{figure}
\centerline{\epsfxsize=3.0truein\epsffile[80 520 320 700]{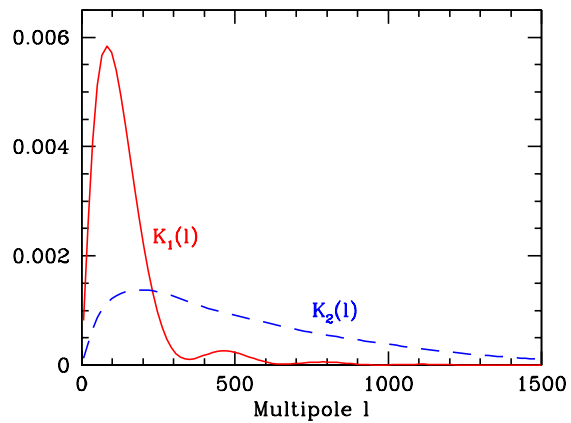}}
\caption{\footnotesize Principal components $K_1(l)$, $K_2(l)$ of the lensing potential $C_\ell^{\phi\phi}$
obtained from CMB measurements to $\ellmax=2000$, as described in \S\ref{sec:pcomp}.
These represent modes in the lensing potential which are constrained by measuring either lensed \{$T$,$E$\} or lensed
$B$-modes respectively.}
\label{fig:eigpp}
\end{figure}

\subsection{Principal Components}
\label{sec:pcomp}

We begin by choosing the parameters of interest to be
fluctuations $p_l$ in the power spectrum of the lenses
around the fiducial model
\be
C_l^{\phi\phi} = (1 + p_l ) C_l^{\phi\phi}|_{\rm fid}\,.
\label{eq:deltaclpp}
\ee
With these parameters in the Fisher matrix of Eq.~(\ref{eq:Fdef}), the
covariance matrix
\be
({\bf C})^{ll'} = ({\bf F}^{-1})^{ll'}
\ee
can be interpreted as that of the measurements of
$C_l^{\phi\phi}$ under the assumption that the parameters that control
the unlensed CMB are fixed.

The principal components or eigenvectors of this covariance matrix determine the best
constrained linear combinations of $C_{l}^{\phi\phi}$.  
We find that if lensing $B$-modes are not observed, the covariance ${\bf C}^{ll'}$ is
dominated by one well-constrained component $K_1(l)$, which we show in Figure \ref{fig:eigpp}.
Equivalently, this means that only one observable in $C_l^{\phi\phi}$ is constrained by
lensed \{$T,E$\} power spectra:
\be
\Theta_1 \eqdef \sum_l {C_l^{\phi\phi} \over C_l^{\phi\phi}|_{\rm fid}}  K_1(l).
\label{eq:Th1def}
\ee
The power spectrum $C_l^{\phi\phi}|_{\rm fid}$ of the fiducial model is scaled out of the
weights $K_1(l)$, such that deviations from $\Theta_1=1$ represent the fractional change
in the weighted amplitude of the power.  Hence the normalization is chosen such that
$\sum_l K_1(l)=1$.
 
On the other hand, if we make the artificial assumption that lensing $B$-modes are observed but
lensed  $\{T,E\}$ are not, then we find that the covariance is dominated by a single broad
component $K_2$, which peaks at $l\sim 200$ and includes a wide range of $l$.
Lensed $B$-mode measurements therefore constrain a second observable  
\be
\Theta_2 \eqdef \sum_l {C_l^{\phi\phi}  \over C_l^{\phi\phi}|_{\rm fid}} K_2(l) \,.
\label{eq:Th2def}
\ee

The principal components $K_i(l)$ were computed assuming cosmic variance limited
CMB measurements to $\ellmax=2000$; however, the shape of the eigenmodes
remains nearly the same if $\ellmax$ is lowered, or if a white noise power spectrum is
used in place of a cutoff in $l$.
Therefore, the observables $\Theta_i$ provide a parameter-independent representation
of the information in the lensed CMB regardless of the noise characteristics.
A caveat to this statement is that we never consider CMB multipoles beyond $\ellmax=2000$
in this paper; relaxing this assumption may permit additional modes in the lensing potential
to be constrained.

For $BB$, the higher principal components are not completely negligible; the
second-best constrained component has a variance which is worse than $K_2(l)$ by
a factor of 7.
We have found that constraints from higher components can almost always be neglected in
parameter forecasts, but can have some impact on degenerate directions 
involving curvature for a measurement
of lensing $B$-modes which is close to all-sky cosmic variance limited.
In the rest of the paper, we will ignore higher components from lensed $B$-modes.

The structure of the two eigenmodes is related to the lensing kernels of
Eq.~(\ref{eq:kerndef}).   Given that power in the deflection angles peaks at $l_{1} <100$ 
in the fiducial model, lensing mainly acts as a convolution kernel of width $l_{1}$
on the high $l$  CMB power spectrum.    The \{$T,E$\} kernels share a similar structure 
since
the angle between the lensed and unlensed ${\bf l}$ is of order $l_{1}/l$.  The $B$ kernel
is weighted toward higher $l_{1}$ for the same reason.  Likewise the dominance of a single
mode in \{$T,E$\} reflects the tight range in $l_{1}$ of the convolution compared with typical
structure in the unlensed power spectra.

\begin{figure}
\centerline{\epsfxsize=3.0truein\epsffile{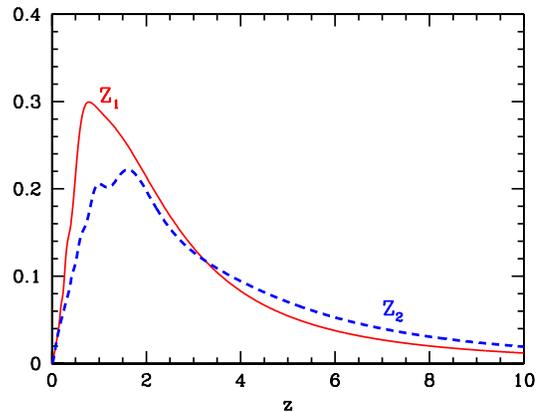}}
\caption{\footnotesize Redshift sensitivity of the lensing observables $\Theta_i$  near the
fiducial model.  To good approximation the observables constrain the amplitude of $C_{l_i}^{\phi\phi}$
around multipoles near the median of the eignmodes of Fig.~\ref{fig:eigpp}, $l_{K1}=114$, $l_{K2}=440$.  The
redshift sensitivities $Z_i$ at these multipoles
(see Eq.~(\ref{eq:redshiftkernel})) are plotted for the fiducial
model. }
\label{fig:redshift}
\end{figure}

\subsection{Parameter Sensitivity}
\label{sec:redshift}

Next to understand how sensitivity to these eigenmodes translates into cosmological parameters,
let us examine their construction in both the multipole and redshift direction.   The change in the
observables due to cosmological parameters can be derived from Eqs.~(\ref{eq:Th1def})-(\ref{eq:Th2def}) once
the change in $C_l^{\phi\phi}$ is known. 

\begin{figure}
\centerline{\epsfxsize=3.0truein\epsffile[80 520 320 700]{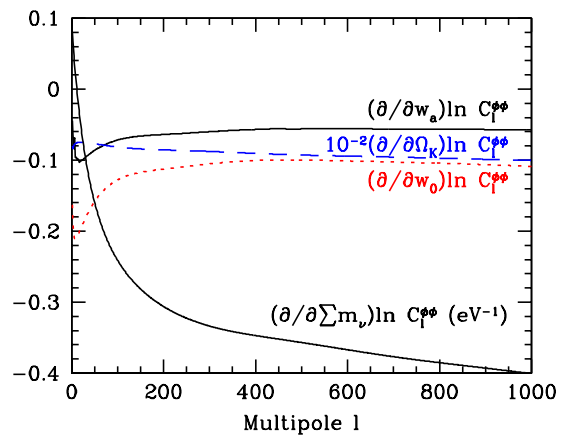}}
\caption{\footnotesize Derivatives of $C_l^{\phi\phi}$ with respect to the parameters $\wn$, $w_0$, $w_a$, 
and $\Omega_K$, illustrating the different $l$ dependence.
As in Tab.~\ref{tab:theta}, the derivatives are taken adjusting $\Omega_{\rm DE}$ to hold $l_A$ fixed.}
\label{fig:derivatives}
\end{figure}

 In Fig.~\ref{fig:derivatives} we plot the derivatives $\partial C_l^{\phi\phi}/\partial p_\alpha$ for several 
cosmological parameters $p_\alpha$.  The corresponding derivatives of the observables are given 
 in Tab.~\ref{tab:theta}.
Since the acoustic peaks constrain 
\begin{equation}
l_A \eqdef {\pi D_A(D_s)/s_s} \,,
\end{equation}
where $s_s$ is the sound horizon at recombination $D_s$, we take these derivatives
at fixed $l_A$ (by adjusting $\Omega_{\rm DE}$).  They then quantify the
additional sensitivity to cosmological parameters introduced by lensing.

  Notice
that the derivatives of the power spectra are quite flat compared across the multipoles where
the two principal components have support  (see Fig.~\ref{fig:eigpp}).   Hence the
sensitivity of the observables to most parameters can be accurately determined from
the sensitivity of the power spectra at the median multipoles $l_{Ki}$ of the principal
components, defined by $\sum_{i=1}^{l_{Ki}} K_{i}(l) = 1/2$:
 $l_{K1}=114$ and $l_{K2}=440$. 

Next, to understand the relative sensitivities to different parameters, consider the fact
that $C_l^{\phi\phi}$ is determined by a projection of the matter power 
spectrum with a well-defined redshift sensitivity.   
In Fig.~\ref{fig:redshift}, we plot this sensitivity $Z_{i}(z)$,  where
\be
C_{l_{K_{i}}}^{\phi\phi} =  C_{l_{K_{i}}}^{\phi\phi}|_{\rm fid} \int dz Z_{i}(z) \,.
\label{eq:redshiftkernel}
\ee
These weights are calculated under
the Limber approximation (see e.g.~\cite{Hu00b}). 
In the fiducial model $\int dz Z_{i}=1$, so that fluctuations in $Z_i$ determine 
fluctuations in the observables as   
\be
\delta \Theta_{i} \approx {\delta C_{l_{Ki}}^{\phi\phi} \over  C_{l_{Ki}}^{\phi\phi}|_{\rm fid} }
= \int dz \, \delta Z_{i}(z)\,.
\ee
We expect this to be a reasonable approximation since ${\delta 
  C_l^{\phi\phi}}$ is not a rapidly varying function of $l$ (see
Figure \ref{fig:derivatives}). 
 
To make the above considerations more concrete, consider the
sensitivity to changes in the distance $D_{A}$, expansion rate $H$,
growth rate of the gravitational potential $G$ and the shape of the
matter power spectrum 
$\Delta^{2}_{m}= k^{3}P(k)/2\pi^{3}$ at the lens redshift 
\bea
{\delta Z_{i} \over Z_{i}} &=& \Big[ n_i {\delta D_{A}\over D_{A}}-{\delta H \over H} + 2 {\delta G \over G} 
+ 2 {\delta D_{A}(D_{s}-D) \over D_{A}(D_{s}-D )}\Big]\,,\nn\\
n_i &\eqdef&  3 - {d \ln \Delta_{m}^{2}\over d\ln k}\Big|_{k=({l_{Ki}/ D_{A}})} \,.
\label{eq:zsens}
\eea
A typical value for the slope of the power spectrum gives $n_i \sim 1$.

CMB lensing is sensitive mainly to
high redshift changes in the amount of lensing and correspondingly 
 $\Theta_{1}$ has a median
redshift of $z \sim 2$ and $\Theta_{2}$, $z\sim 3$. 
The observables best constrain cosmological parameters that change the geometry and growth in 
a coherent fashion at these redshifts. 
For example, a negative spatial curvature decreases the angular diameter distance
to the lens redshifts given the fixed $D_s$ and also decreases the growth rate.  Both of these
effects persist to $z\sim 2-3$ and hence lensing is highly sensitive to the curvature.   
Likewise massive neutrinos slow the growth from the time they become non-relativistic
near recombination.  They also create a substantial change at high redshift.

For the dark energy, we take  its equation of state to be parameterized by 
\{$w_0$, $w_a$\}  according to Eq.~(\ref{eq:wa}).  These parameters
suffer in sensitivity in that the changes they induce
on the geometric and growth parameters are dominant at $z<1$.  Furthermore their effects on
the expansion rate, distance and growth tend to cancel in the observables (see \cite{Hu04b} for a more extended discussion).  Note that this a feature of the specific parameter set chosen 
 and may not apply to all dark energy models.  The principal component
technique allows a parameter independent way of quantifying the lensing information.

The geometric and growth parameters are not the only ones that affect the observables.
Since the observables are directly related to the power spectrum of the lenses any parameter
that alters it also alters the observables causing parameter
degeneracies (see \S \ref{sec:exfisher}).
The most important of these are $\Omega_c h^2$
and the amplitude of the initial
power spectrum $\delta_\zeta$.  The sensitivity of the observables to these parameters 
are given in Tab.~\ref{tab:theta}.
For completeness, we also
 give the sensitivity to $l_A$, though
errors on this parameter from the acoustic peaks will be negligible for lensing purposes.

\begin{table}
\begin{center}
\begin{tabular}{|c|c|c|}
\hline                                         &   $\Theta_1$   &   $\Theta_2$   \\  \hline
$\partial\Theta_i/\partial(\wn)$    &    -0.24       &    -0.34       \\
$\partial\Theta_i/\partial w_0$                  &    -0.14       &    -0.12       \\
$\partial\Theta_i/\partial w_a$                &    -0.072      &    -0.061      \\
$\partial\Theta_i/\partial \Omega_K$           &    -8.24       &    -9.17       \\  \hline
$\partial\Theta_i/\partial (\wm)$     &     17.0       &     24.7       \\
$\partial\Theta_i/\partial \ldz$               &     2.00       &     2.09       \\
$\partial\Theta_i/\partial \ln l_A$ &  2.37 & 2.99 \\   \hline
\end{tabular}
\end{center}
\caption{\footnotesize Derivatives of the observables $\Theta_1$, $\Theta_2$ with respect to parameters of interest (top) and nuisance parameters (bottom).  In all rows except the last, the derivatives are taken
adjusting $\Omega_{\rm DE}$ to hold $l_A$ fixed.   Units for $\wn$ are eV.}
\label{tab:theta}
\end{table}

\subsection{Sample and Noise Variance}
\label{sec:lensvariance}

The  sample and noise variance of the lensing observables determines
the errors on cosmological parameters.  Furthermore, this characterization of the errors
illuminates the origin of the non-Gaussian bandpower covariance.

\begin{figure}
\centerline{\epsfxsize=3.0truein\epsffile[60 360 320 700]{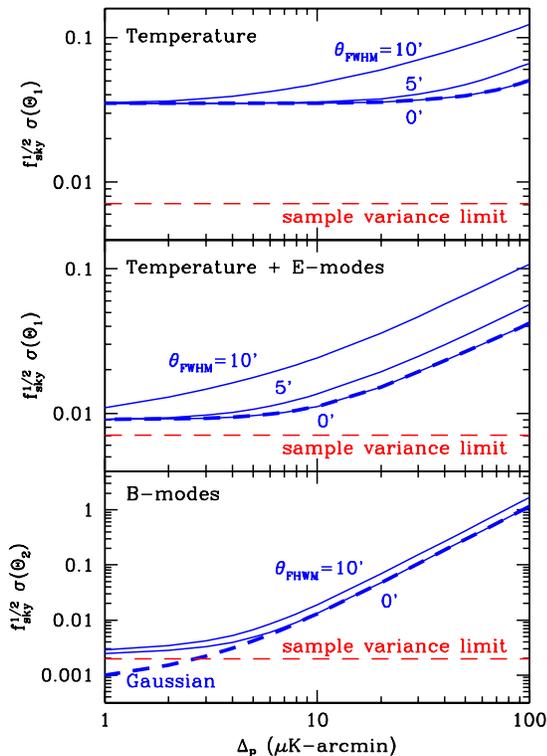}}
\caption{\footnotesize Uncertainty on $\Theta_1$ from lensed $T$ alone (top) and lensed \{$T,E$\} (middle),
and uncertainty on $\Theta_2$ from lensed $B$ (bottom), for varying beam size and noise level
$\Delta_P$.  We assume $\Delta_T$ is given by $\Delta_P/\sqrt{2}$ throughout, including the top panel.
Only multipoles below $l_{\rm max}=2000$ are included.
For the zero beam cases, we also show the uncertainties that would be obtained if Gaussian statistics were
falsely assumed (dashed).  The impact of non-Gaussian contributions is negligible in \{$T,E$\} but
significant for $B$.
The horizontal lines are sample variance limits given by Eq.~(\ref{eq:svlimit}).}
\label{fig:sigmatheta}
\end{figure}

In Figure \ref{fig:sigmatheta}, we show the $1\sigma$ errors on $\Theta_i$ which are obtained
for different instrumental sensitivities.  Combined with the sensitivity of $\Theta_i$ to
cosmological parameters, these results may be used to forecast 
parameter errors (see \S \ref{sec:exfisher}).  

As the instrumental noise goes to zero, the errors saturate at the combined sample variance
limits of the lenses and unlensed CMB.   To understand the relationship between sample
variance and non-Gaussian covariance, consider first an ideal noise-free direct measurement
of the lensing potential $\phi$.
From the definition~(\ref{eq:Th1def}), the sample variance limit on a measurement of the
observables arising from the lenses is given by
\bea
\sigma^2_{\rm SV}(\Theta_i) &=& \sum_{l,l'} K_i(l) K_i(l') ({\bf C}_{\rm SV})^{ll'} \nn \\
                        &=& f_{\rm sky}^{-1} \sum_l \frac{2}{2l+1} K_i(l)^2 \,.
\label{eq:svlimit}
\eea
Here we have taken ${\bf C}_{\rm SV}^{ll'} = 2f_{\rm sky}^{-1}\delta_{ll'} /(2l+1)$ under the assumption that
$\phi$ is a Gaussian field.

In Figure \ref{fig:sigmatheta}, the sample variance limits of Eq.~(\ref{eq:svlimit}) are shown as dashed
horizontal lines.  In the limit of zero noise, lensed CMB measurements can measure
$\Theta_1$ and $\Theta_2$ at nearly their sample variance limits.
Direct measurements of the $\phi$ field, e.g.~from CMB lens reconstruction,
do however contain more information than the observables $\Theta_i$.  They are only two of many quantities that can be constructed
with a sample variance limited ${\bf C}_{\rm SV}^{ll'}$.   These other quantities are related
to changes in the shape of $C_l^{\phi\phi}$ and may be useful for breaking degeneracies. 

The two observables do however contain the majority of
the low $l$ information on the amplitude of $C_l^{\phi\phi}$.  To see this, consider the mode
$p_l = A-1$ in  Eq.~(\ref{eq:deltaclpp}).    This parameter can be measured to
\be
\sigma^2_{\rm SV}(A) = \left[ \sum_{l\le l_{\phi{\rm max}}}  {2l+1 \over 2} f_{\rm sky}\right]^{-1} \,.
\label{eq:Alimit}
\ee
The qualitative difference between Eqs.~(\ref{eq:svlimit}) and (\ref{eq:Alimit}) is that the former
is limited by the multipoles with the largest weighted sample variance. The latter is
limited by the smallest and hence determined by the  cutoff $l_{\phi{\rm max}}$. 
In the fiducial model $\sigma^2_{\rm SV}(A) = \sigma_{\rm SV}^2(\Theta_1)$ at $l_{\phi{\rm max}}=198$
and $\sigma^2_{\rm SV}(A) = \sigma_{\rm SV}^2(\Theta_2)$ at  $l_{\phi{\rm max}}=705$.  Given
that CMB power spectrum measurements out to $l_{\rm max}=2000$ capture nearly all
of the information on the two observables, they also
capture essentially all of the information on the amplitude of the lens power spectra
near the median of the weights of the eigenmodes.

Finally, if Gaussian statistics are falsely assumed for $BB$, then one would conclude
that a high sensitivity measurement of CMB $B$-modes constrains the lensing observable $\Theta_2$
better than the sample variance limit (see Fig.~\ref{fig:sigmatheta}).
From this perspective, one can gain intuition into why non-Gaussianity is significant for lensed
$B$-modes.  If the $B$-modes were perfectly Gaussian, then the overall amplitude of the lensing could be
constrained to within the sample variance of the smallest scale fluctuations in the CMB.
In reality, the amplitude of the lensing is limited by the sample variance of the lenses near the
median of the eigenvectors, i.e.~on degree
scales.  Band powers below this scale covary in amplitude since the induced $B$-modes share
the same lens fluctuations.   This covariance becomes noticeable when the intrinsic sample variance
of the $E$-modes becomes subdominant due to binning.

The principal component analysis also shows why non-Gaussianity is not a significant limitation for
 \{$T, E$\}.
The Gaussian errors on $\Theta_1$ from observations of lensed \{$T,E$\} (Fig. \ref{fig:sigmatheta}) never
exceed the sample variance limit.  This is because the sample variance of the high $l$ unlensed
$\{ T, E \}$ fields still dominate the measurement of the small fractional changes induced by lensing.
The $B$ field does not suffer from this problem in that at high $l$ it is completely generated by lensing.

\section{Parameter Case Studies}
\label{sec:exfisher}

We now study parameter constraints from the lensed CMB, with non-Gaussian contributions to the
power spectrum covariance included.
In general, parameters which affect the high-redshift universe
or the angular diameter distance to last scattering are well measured even without lensing.  Lensing 
mainly helps in breaking degeneracies that leave the observables at recombination fixed.  
We illustrate this degeneracy breaking with massive neutrinos \{$\wn$\}, a constant dark energy equation of
state \{$w$\}, equation of state evolution \{$w_{0}$,$w_a$\}, and spatial curvature \{$\Omega_K$\}.
Given that the lensed CMB adds two new observables as described in the previous section,
we will study these additional parameters two at a time.
We start by examining the \{$\wn,w$\} case extensively, 
as an illustration of the impact of non-Gaussianity,
as well as a worked example of the use of the lensing observables $\Theta_i$.

\begin{table}
\begin{center}
\begin{tabular}{|c|c|c|c|c|c|}
\hline &  $\nu$  & $\theta_{\rm FWHM}$   & $\Delta_T$   &  $\Delta_P$   & $\fsky$ \\  \hline
Planck & 100 GHz  &  $9.2'$  &  51   &       -- & 0.8 \\
       & 143 GHz  &  $7.1'$  &  43  &  78   & 0.8  \\
       & 217 GHz  &  $5.0'$  &  65     &  135 & 0.8    \\  \hline
Deep$_{10\%}$  & -- & 1.0' & 1.00 & 1.41 & 0.1 \\ \hline
\end{tabular}
\end{center}
\caption{\footnotesize Assumed experimental specifications for the reference survey in \S\ref{sec:exfisher},
consisting of Planck and a deep polarization field with 10\% of sky.
The noise parameters $\Delta_T$ and $\Delta_P$ are given in units of
$\mu$K-arcmin.   For the combined reference survey, $\sigma(\Theta_1)=0.025$ and $\sigma(\Theta_2)=0.008$.}
\label{tab:planck}
\end{table}

Throughout this section, we will consider a reference survey (Tab.~\ref{tab:planck}) which consists of Planck
measurements on most of the sky, combined with a smaller deep survey which measures lensing of
the temperature and polarization with
higher signal-to-noise.
(We assume that the deep survey area is a subset of the Planck survey area.)

\subsection{Neutrinos and constant equation of state}

We consider here the constraints in the \{$\wn,w$\} plane at fixed $\Omega_{K}$
employing a direct Fisher matrix forecast
in the parameter space and through the intermediary measurement of the lensing observables,
first by adding its information to the unlensed spectra in the full parameter space and then by
pre-marginalizing parameters that control the high redshift physics at recombination.

\subsubsection{Direct forecasts}
\label{sec:direct}

\begin{figure*}
\centerline{\epsfxsize=7.0truein\epsffile[60 360 550 700]{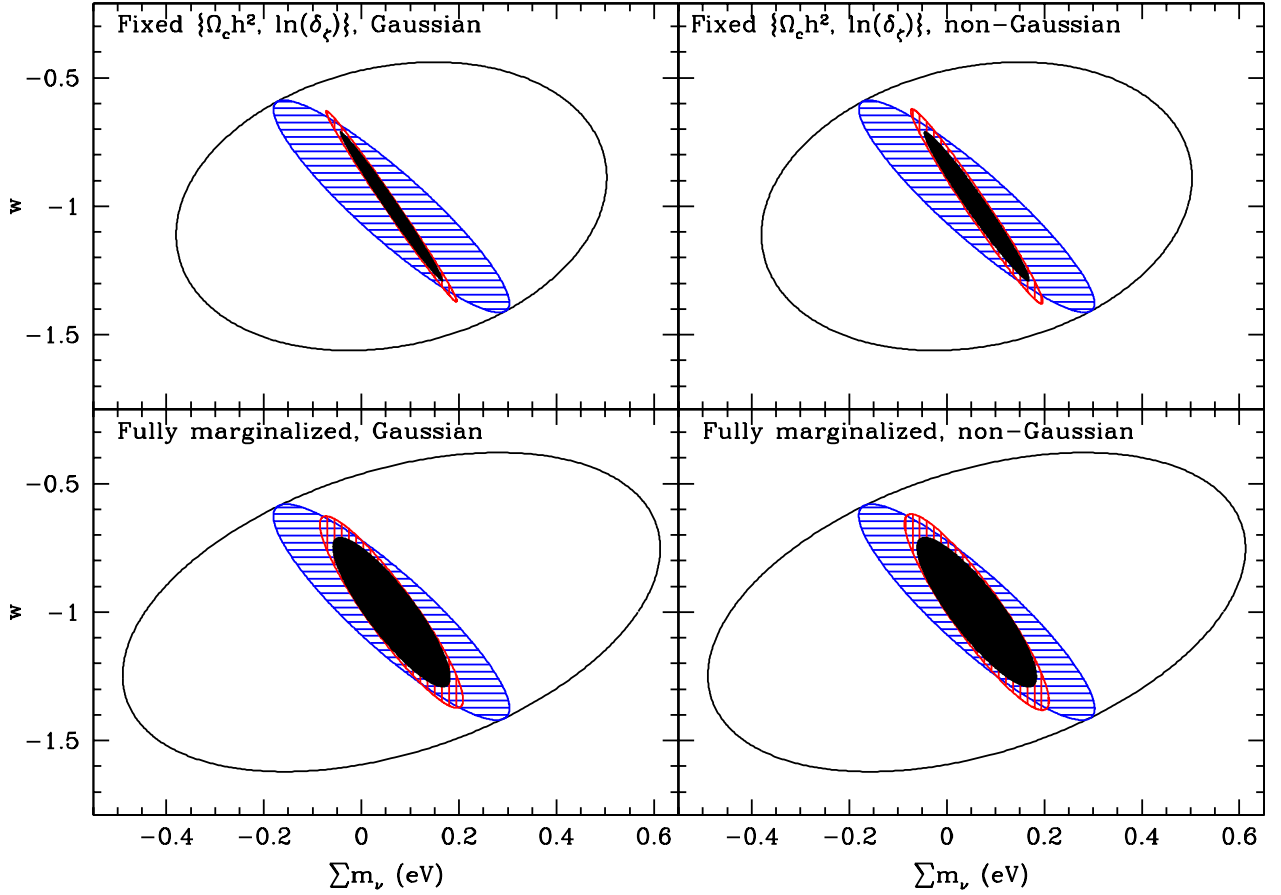}}
\caption{\footnotesize Degeneracy breaking in the $\wn-w$ plane from CMB lensing, for the reference
survey in Tab.~\ref{tab:planck}.  Ellipses here and throughout are plotted at $\Delta\chi^2=1$ and
not 68\% CL.
In each panel, the surrounding ellipse represents parameter constraints from unlensed \{$T,E$\}, the
blue/horizontally-shaded ellipse represents constraints from lensed \{$T,E$\}, the red/vertically-shaded ellipse
represents constraints from unlensed \{$T,E$\} + lensed $B$, and the inner solid ellipse represents constraints
from lensed \{$T,E,B$\}.
In the top panels, we show Gaussian (top left) and non-Gaussian (top right) constraints with
\{$\wm,\ldz$\} held fixed, and the remaining parameters
 \{$\Omega_b h^2, \Omega_{\rm DE}, \tau, n_s, r$\} marginalized.
In the bottom panels, we show Gaussian (bottom left) and non-Gaussian (bottom right) constraints with
all parameters marginalized.}
\label{fig:omnh2_w}
\end{figure*}

In Figure \ref{fig:omnh2_w}, left panels, 
we show the constraints falsely assuming Gaussian statistics.
For illustrative purposes, we first show the errors with the parameters 
\{$\wm$, $\ldz$\}, fixed to their fiducial
values (top left), but  with  the remaining parameters \{$\Omega_{\rm DE}$, $\Omega_b h^2$,
 $\tau$, $n_s$, $r$\} marginalized.  
As we shall see, the former two parameters affect the lensing observables and cause
degeneracies with the parameters of interest.

The large surrounding ellipse in the top left panel
represents the double angular diameter distance degeneracy expected from the unlensed CMB.
Adding information from lensed \{$T,E$\} (horizontally shaded) constrains a slightly different combination of
degenerate parameters than adding information from lensed $B$ (vertically shaded).
This is because changes in the two parameters \{$\wn,w$\} affect $C_l^{\phi\phi}$ differently in $l$ 
(Fig.~\ref{fig:derivatives}) and hence the two lensing observables described in \S\ref{sec:deobs}.

We next consider the effect of including the non-Gaussian covariance, in the top right panel of Fig. \ref{fig:omnh2_w}.
Compared to Gaussian uncertainties, the constraints from lensed \{$T,E$\} are unaffected
(as expected from \S\ref{sec:psest}), while constraints from lensed $B$ are degraded for the
best constrained combination of the two parameters.   Note that because we have considered
a nearly degenerate two parameter space, even here the effect of non-Gaussianity is hidden
from the errors on a single parameter marginalized over the other.

However, the discussion so far has assumed perfect priors for the parameters \{$\wm,\ldz$\}.
If these are marginalized with only information from the reference survey,
the effect of non-Gaussianity is overwhelmed by the
effect of marginalizing (see Fig.~\ref{fig:omnh2_w}, bottom panels).
 This is true even if $B$ were the only source
of lensing information or if there were only one additional parameter.

Nonetheless the effect of non-Gaussianity in the full parameter space
is not negligible; it does enlarge the volume which is allowed.
The enlargement occurs along a direction which is a combination of several parameters
\{$\wm,\ldz,w,\wn$\}.  Though hidden by marginalization, this degradation can be
exposed if there are external prior measurements of other combinations
of these parameters.  It can also reduce the apparent goodness of fit to the best-fit model (see Appendix \ref{sec:chisq}).  
Finally for different reference surveys, e.g.~a smaller but deeper survey,
a Gaussian assumption can again give a misleading answer on the utility of the $B$-modes
(see \S \ref{sec:applications}).

These results can be reproduced and understood using the lensing observables $\Theta_i$ as
we shall now see.  

\begin{figure}
\centerline{\epsfxsize=3.0truein\epsffile[60 510 320 700]{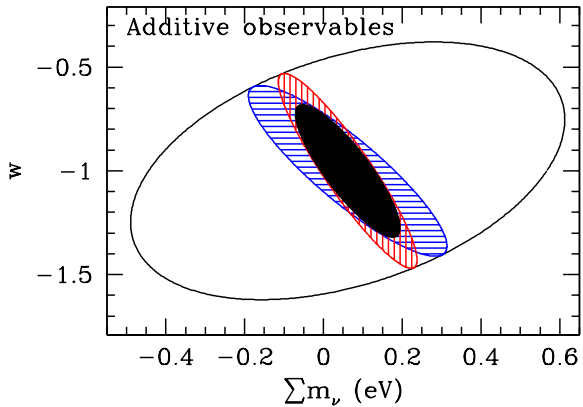}}
\caption{\footnotesize Forecasted errors in \{$\wn,w$\} as in
Fig.~\ref{fig:omnh2_w} (lower right panel, non-Gaussian, fully marginalized) 
but with the additive lensing observables prescription from \S\ref{sec:additive}.
Note the excellent agreement between the direct and observables approaches.}
\label{fig:additive}
\end{figure}

\subsubsection{Additive observables forecasts}
\label{sec:additive}

The lensing observables provide a general framework for forecasting the additional constraints
supplied by lensing in any parameter space.  We can use the direct forecasts from the preceding
section as a basis for comparison with the following construction:
\begin{enumerate}
\item From the parameters of the survey under consideration (sky coverage, noise, and beam),
compute uncertainties (see Fig.~\ref{fig:sigmatheta}) on the observables $\Theta_i$
and assume they are independent, obtaining a 2$\times$2 covariance matrix:
\be
\Cov_{\meas}(\Theta_i,\Theta_j) = \left( \begin{array}{cc}
  \sigma^2(\Theta_1)  &     0                  \\
          0           &  \sigma^2(\Theta_2) 
\end{array} \right)\,.
\label{eq:covmeas}
\ee
\item Transform these parameter errors into a Fisher matrix in the desired parameter space
with
\be
F_{\alpha\beta}^{\rm lens} = (\partial_\alpha \Theta^i) \Cov(\Theta_i,\Theta_j)^{-1} (\partial_\beta \Theta^j) \,
\label{eq:F2by2}
\ee
and add this to the Fisher matrix of the unlensed CMB or any external data set. 
\end{enumerate}

 Some
of the more common parameter derivatives for use in Eq.~(\ref{eq:F2by2})
are given in Tab.~\ref{tab:theta}.  Others can
be approximated by evaluating $C_l^{\phi\phi}$ at the median multipole. 
An  advantage of the observables scheme  is that it automatically includes the effect of
non-Gaussianity without reference to the bandpower covariance matrix.  
One simply uses the non-Gaussian errors from Fig.~\ref{fig:sigmatheta}.

The main caveat to the observables prescription is that it
 implicitly assumes that lensing is an additive source of information. It should
not be used to forecast parameters for which lensing destroys information such as tensors
or features in the initial power spectrum.  
In the particular case of the tensor-to-scalar ratio $r$, we find that non-Gaussianity in the lensed CMB is always
negligible when forecasting uncertainties, even when low-redshift parameters are marginalized.

We show the $w-\sum m_\nu$ example in Fig.~\ref{fig:additive}.  Note the excellent agreement
with the direct Fisher calculation in Fig.~\ref{fig:omnh2_w} (lower right panel) in all respects.

\subsubsection{Nuisance-marginalized observables}
\label{sec:premarg}

This prescription can be further simplified so as not to require computation of the unlensed CMB
Fisher matrix or external information
  for forecasting parameter errors which
rely mainly on the lensing information.
In this case the extra information is utilized to remove the parameter degeneracy 
from $\wm$ and $\ldz$ in the observables $\Theta_{i}$.  
These parameters can be premarginalized and
 dropped from the Fisher matrix constructed from $\Theta_{i}$.

Operationally, to pre-marginalize the nuisance parameters,
we add the ``nuisance errors''
\bea
\Cov_{\rm nuis}(\Theta_i,\Theta_j) &\eqdef&
\frac{\partial\Theta_{i}}{\partial\wm} \frac{\partial\Theta_{j}}{\partial\wm} \sigma^{2}(\wm)  \label{eq:covnuis}  \\
&& \qquad +\, \frac{\partial\Theta_{i}}{\partial\ldz} \frac{\partial\Theta_{j}}{\partial\ldz} \sigma^{2}(\ldz)  \nn
\eea
to the ``measurement errors'' defined by Eq. (\ref{eq:covmeas}),
to get an effective covariance matrix $\Cov_{\tot} = (\Cov_{\meas} + \Cov_{\rm nuis})$.
We then use $\Cov_{\tot}$ instead of $\Cov_{\meas}$ in Eq.~(\ref{eq:F2by2}),
to obtain a Fisher matrix in a parameter space where  \{$\wm,\ldz$\} is excluded.

In~(\ref{eq:covnuis}), we have neglected the correlations between $\wm$ and $\ldz$ but these provide
a negligible effect if the source of information is internal to the CMB.  Note that 
$\Cov_{\tot}$ will typically show a high degree of correlation between the two observables
since the increase in the effective errors is along directions that are degenerate with the
two parameters.  For parameters that do not induce a degenerate change, this correlation
reflects extra information (see \S \ref{sec:optimization}).

\begin{table}
\begin{center}
\begin{tabular}{|c|c|c|c|c|}
\hline                                  &    WMAP3  &  Planck  &  Reference  &  Ideal   \\   \hline
$\sigma(\Theta_{1})$          &      -    &  0.050   &    0.025    &  0.0089  \\
$\sigma (\Theta_{2})$          &      -    &    -     &    0.008    &  0.0023  \\   \hline
$\sigma(\wm)$                            &    0.01   &  0.0011  &    0.0009   &  0.0005  \\
$\sigma(\ldz)$                           &    0.03   &  0.0045  &    0.0040   &  0.0023  \\
$\sigma_{\rm nuis}(\Theta_1)$            &    0.18   &  0.020   &    0.017    &  0.0094  \\
$\sigma_{\rm nuis}(\Theta_2)$            &    0.25   &  0.028   &    0.023    &  0.011   \\
$\Corr_{\rm nuis}(\Theta_1,\Theta_2)$    &    0.99   &  0.99    &     0.99    &  0.99    \\   \hline
$\sigma_{\tot}(\Theta_1)$             &      -    &  0.054   &    0.030    &  0.013   \\
$\sigma_{\tot}(\Theta_2)$             &      -    &     -    &    0.025    &  0.013   \\
$\Corr_{\tot}(\Theta_1,\Theta_2)$     &      -    &     -    &     0.52    &  0.71    \\ \hline
\end{tabular}
\end{center}
\caption{\footnotesize 
Uncertainties in lensing observables from third-year WMAP \cite{Speetal06},
Planck alone (Tab.~\ref{tab:planck}, top), the reference experiment consisting
of both parts of Tab.~\ref{tab:planck}, and an ideal survey which is all-sky
cosmic variance limited in temperature and polarization to $\ellmax=2000$.
{\it Top:} Measurement errors on the lensing observables \{$\Theta_1,\Theta_2$\} computed from
the raw sensitivity to lensed \{$T,E$\} and $B$ respectively, as in Fig.~\ref{fig:sigmatheta}.
{\it Middle:} ``Nuisance'' errors of Eq.~(\ref{eq:covnuis}) on lensing observables, computed by
propagating each survey's unlensed uncertainties on \{$\wm,\ldz$\} (also shown) using derivatives
from Tab.~\ref{tab:theta}.
{\it Bottom:} Total errors on lensing observables, computed by adding 2$\times$2 covariance
matrices ($\Cov_{\tot} = \Cov_{\meas} + \Cov_{\rm nuis}$).  As described in the text, these
are the effective errors on lensing observables when constraining low-redshift parameters with
nuisance parameters \{$\wm,\ldz$\} premarginalized.  }
\label{tab:nuisance}
\end{table}

\begin{figure}
\centerline{\epsfxsize=3.0truein\epsffile[60 360 320 700]{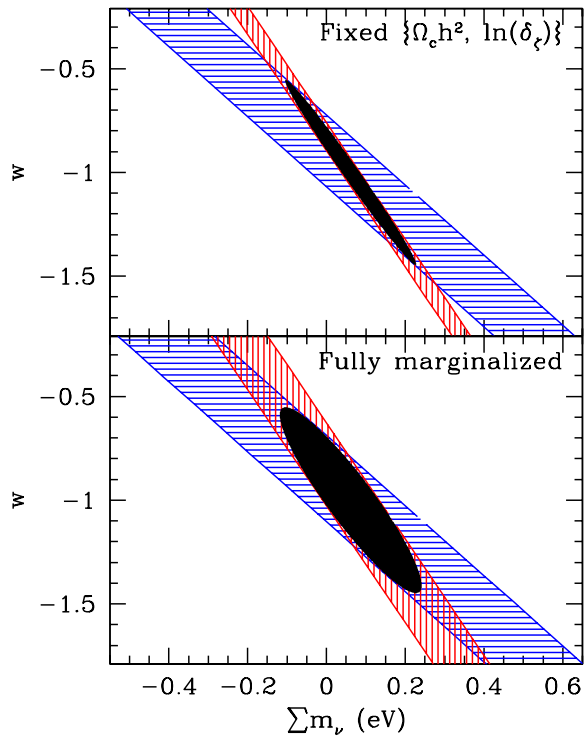}}
\caption{\footnotesize Forecasted errors in \{$\wn,w$\} as in
Fig.~\ref{fig:omnh2_w} (right panels, non-Gaussian) 
but with the premarginalized-observables prescription from \S\ref{sec:premarg}.
The blue/horizontally-shaded regions show only the $\Theta_1$ constraint from lensed \{$T,E$\},
the red/vertically-shaded regions show only the $\Theta_2$ constraint from lensed $B$, and
the solid regions show the combined constraint from both observables.  The top panel shows the
case where \{$\wm, \ldz$\} are completely fixed by the external prior; the bottom panel shows
the case where the parameters are internally determined by the reference survey itself.}
\label{fig:approx_nw}
\end{figure}

Let us illustrate this  pre-marginalization scheme for  \{$\wn, w$\} and the reference survey. 
For comparison with Fig.~\ref{fig:omnh2_w}, we will compute errors  
with priors on \{$\wm,\ldz$\} that fix them completely, and priors that are derived from determinations
internal to the reference survey (see Tab.~\ref{tab:nuisance}).  
The former case also corresponds to using
$\Cov_{\meas}$ (Eq.~(\ref{eq:covmeas})) in place of $\Cov_{\tot}$.
Combined with the parameter derivatives from Tab.~\ref{tab:theta},
the result of this procedure is shown in Fig.~\ref{fig:approx_nw}.

The constraints agree well with the direct Fisher calculation in the right
panels of Fig.~\ref{fig:omnh2_w} along the best constrained direction.
For example, with $\sum m_\nu$ fixed, the uncertainty $\sigma(w)$ is
0.150 from the direct Fisher calculation (\S\ref{sec:direct}), 0.170 from
the additive-observables prescription (\S\ref{sec:additive}), and 0.178 from
the premarginalized-observables prescription described in this section.
In practice, the main difference between the last two is that in the premarginalized-observables
prescription, the constraints computed are from lensing alone; information from the unlensed
CMB which helps constrain intermediate-redshift parameters (such as ISW) is not included.

A pedagogical advantage of pre-marginalization is that it clarifies the limiting source of uncertainty.
For example, in our reference survey, $\sigma_{\rm nuis}(\Theta_2)$ is substantially 
larger than $\sigma_{\meas}(\Theta_{2})$.
The utility of the lensing observable is therefore limited by the nuisance parameters and not by the sample or noise
variance on lensing.   This also explains the negligible impact of non-Gaussian sample
variance on marginalized parameter errors in Fig.~\ref{fig:omnh2_w}.

\subsection{Evolution of equation of state}

We now consider CMB constraints on a time-dependent dark energy equation of state through
\{$w_{0},w_{a}$\} [see Eq.~(\ref{eq:wa})].
For purposes of this subsection, we assume that both the neutrino mass and spatial curvature
are fixed and marginalize
\{$\Omega_{\rm DE}$, $\Omega_b h^2$, $\wm$,  $\tau$, $n_s$, $r$\}.

\begin{figure}
\centerline{\epsfxsize=3.0truein\epsffile[60 360 320 700]{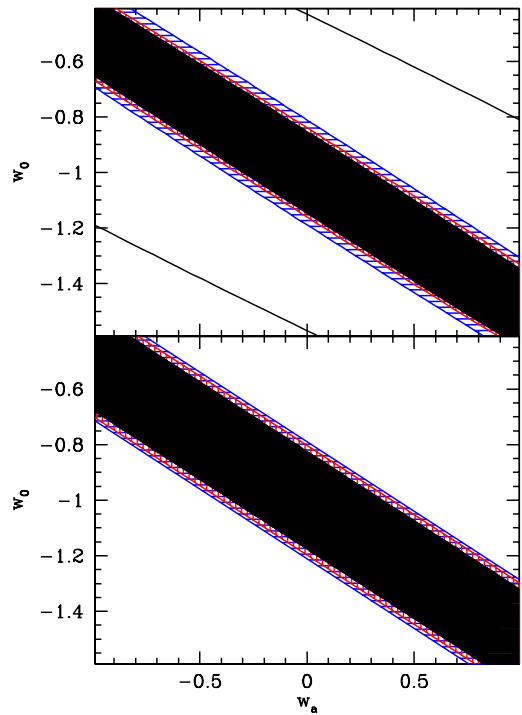}}
\caption{\footnotesize Joint uncertainties on dark energy parameters \{$w_a,w_0$\}, for the reference survey
in Tab.~\ref{tab:planck}, with \{$\wn,\Omega_K$\} fixed to their
 fiducial values and all other parameters marginalized.
The top panel shows the direct Fisher matrix calculation; the bottom panel shows errors calculated using the 
premarginalized-observables scheme (\S\ref{sec:premarg}).
In this parameter space, the lensing observables are degenerate; the $\Theta_1$ constraints (blue/horizontally-shaded),
the $\Theta_2$ constraints (red/vertically-shaded), and the combined constraints (solid) are nearly identical.  Given this nearly perfect degeneracy, the Fisher forecast should not be interpreted
literally along the full extent of the degeneracy. }
\label{fig:wwa}
\end{figure}

In Fig. \ref{fig:wwa}, 
we show errors on $w_{0}$ and $w_a$, for the reference survey of Tab.~\ref{tab:planck},
using the premarginalized observables scheme from \S\ref{sec:premarg}.
In the top panel we show the direct Fisher matrix calculation for comparison.
One parameter combination is constrained by CMB lensing, but the complementary direction is degenerate,
acquiring CMB constraints only from the unlensed CMB through the ISW effect
(c.f.~\cite{AcqBac05}).

The \{$w_0,w_a$\} degeneracy can be seen directly by noting
that the $w_0$ and $w_a$ derivatives, taken at constant angular diameter distance,
of $C_l^{\phi\phi}$ are nearly proportional (Fig. \ref{fig:derivatives}), so that a parameter space direction
exists which preserves both $C_l^{\phi\phi}$ and the unlensed CMB.
This is also seen in Tab.~\ref{tab:theta}, where the $w_0$ and $w_a$ derivatives of \{$\Theta_1,\Theta_2$\}
are nearly proportional.

The degeneracy can be interpreted as the statement that the lensed CMB constrains $w(z)$
mainly around the pivot redshift $z_{p}\sim 1$ with uncertainty $\sigma(w_{p}) = 0.15$
for the reference survey.
The pivot redshift can be interpreted as a combination of the high redshift weight of
CMB lensing discussed in \S \ref{sec:redshift} and the dark energy parameters which change
observables strongly only at $z<1$.  Note that this pivot redshift is substantially higher than
many other cosmological probes and implies that CMB lensing will provide
complementary constraints on the evolution of the equation of state.
We also note that the pivot redshift and $\sigma(w_p)$ will
depend on the underlying fiducial model.

\subsection{Curvature}
\label{sec:curvature}

\begin{figure}
\centerline{\epsfxsize=3.0truein\epsffile[60 360 320 700]{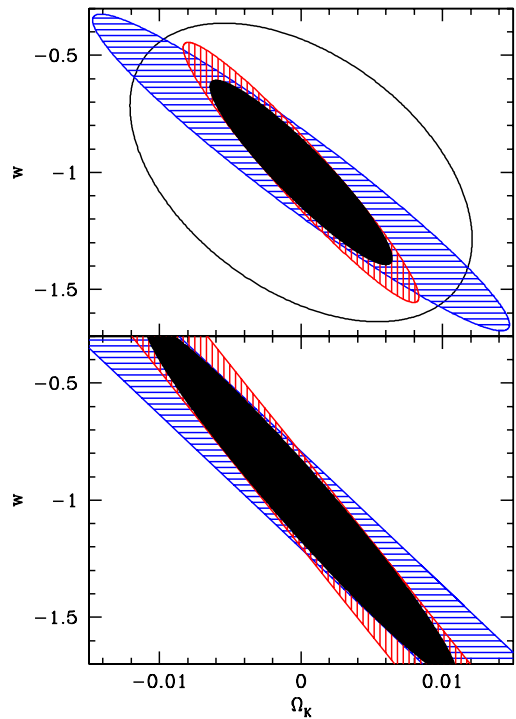}}
\caption{\footnotesize Joint uncertainties on \{$\Omega_K,w_0$\}, for the reference survey of Tab.~\ref{tab:planck},
with \{$\wn,w_a$\} fixed to their fiducial values and all other parameters marginalized.
The top panel shows the direct Fisher matrix calculation; the bottom panel shows errors calculated using the 
premarginalized-observables scheme from \S\ref{sec:premarg}.
As expected, the two agree in the well constrained direction, 
but the second scheme overestimates the errors in the poorly constrained, ISW-limited, direction.}
\label{fig:kw}
\end{figure}

Given the sensitivity of CMB lensing to changes at high redshift, constraints on spatial
curvature are much stronger than those on the dark energy when compared
with low-redshift probes of the expansion history.  In Fig.~\ref{fig:kw}, we
show the reference constraints in the \{$\Omega_{K}, w$\} plane with
\{$\Omega_b h^2, \wm, \Omega_{\rm DE}, \tau, n_s, r$\} marginalized
and \{$\wn,w_a$\} fixed.

The two observables are again nearly degenerate in this plane.   The direction of the
degeneracy is well but not perfectly matched in the observables scheme for the case
of $\Theta_1$.   The reason for this is that the unlensed CMB carries information on the
curvature both from the ISW effect and from the intrinsic sharpness of the
acoustic peaks (see \cite{HuWhi96b}, Fig. 11).  The latter effect comes from the geometrical
projection of $k$-space power to $l$-space through the radial eigenfunctions.  Unlike
lensing this effect smooths (negative curvature) or sharpens (positive curvature) even the
low order peaks as it is associated with the curvature across the last scattering surface
as a whole.  For the high order peaks it destructively interferes with the lensing effect
and mildly violates the assumption that lensing is an independent additive source of
information.   The interference and the rotation it causes
however are small and, if desired, can be accounted for
by a $\sim 20\%$ lowering of $\partial \Theta_1 /\partial \Omega_K$ for surveys
that utilize information near $\ellmax=2000$.

More importantly, the direction that is well constrained has a large $\Omega_K$ and
only a small $w$ component, i.e.~the degeneracy is very steep in the curvature
direction.  The implication is that it takes only a weak external constraint on $w$
to break this degeneracy completely.   When combined with other dark energy
probes, the lensing observables can be thought of as fixing the curvature.  We
explore this use of CMB lensing further in a separate piece \cite{HuHutSmi06}.

\begin{figure}
\centerline{\epsfxsize=3.0truein\epsffile[60 360 320 700]{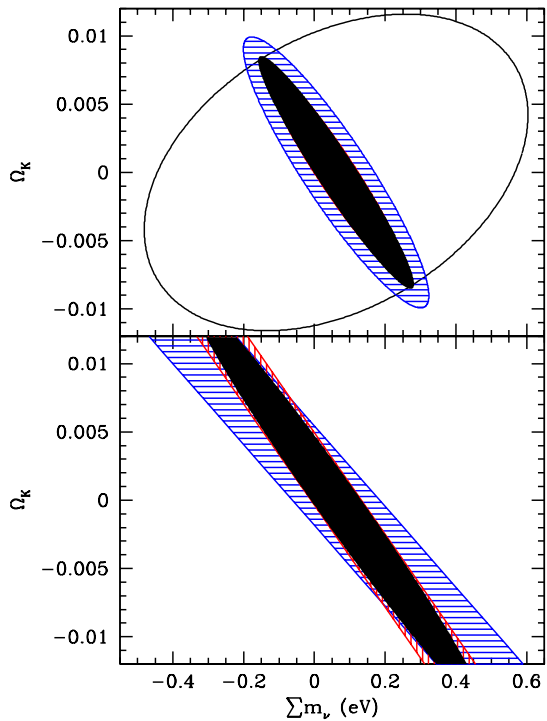}}
\caption{\footnotesize Joint uncertainties on dark energy parameters \{$\wn,\Omega_K$\}
with \{$w_0,w_a$\} fixed to their fiducial values. 
for the reference survey of Tab.~\ref{tab:planck}, using the direct Fisher calculation (top panel)
and the premarginalized-observables scheme (bottom panel).
(In the top panel, the red/vertically-shaded ellipse representing unlensed \{$T,E$\} and lensed $B$
is hidden by the solid ellipse representing lensed \{$T,E,B$\}.)}
\label{fig:nk}
\end{figure}

Even given other probes that break the \{$\Omega_{K}, w$\} degeneracy,
$\Omega_K$ remains nearly degenerate with $\wn$.  In Fig.~\ref{fig:nk}, we show
constraints in this plane with \{$w_0, w_a$\} fixed.   The observables approach again accurately
models the well constrained direction aside from the slight rotation of the $\Theta_1$ constraint 
which makes curvature nearly perfectly degenerate with neutrinos. 
Breaking this degeneracy externally will require independent probes with limits
of $\sigma(\Omega_K) \ll 0.005$ (perhaps with measurements of 
the angular diameter distance to $z \sim 3$ \cite{Kno05}) or
$\sigma(\sum m_\nu) \ll 0.2$eV.

\section{Survey Optimization}
\label{sec:applications}

In the previous section, we quantified the information supplied by CMB lensing
in the context of a specific reference survey.
We conclude this paper by considering how surveys can optimize the
extraction of cosmological information from lensed power spectra.

\subsection{$\wm$, $\tau$,  and $\ellmax$}
\label{sec:wmtau}

We have seen in \S\ref{sec:deobs} that imperfect knowledge of \{$\wm$, $\ldz$\} results in nuisance errors
which limit the ability to extract cosmological information from lensing.
The nuisance errors $\sigma_{\rm nuis}(\Theta_i)$ represent a floor below
 which improving sensitivity to each lensing observable individually does not improve cosmological
parameter constraints.
More precisely, improving $\sigma(\Theta_i)$ beyond this level will correlate low-redshift
parameters to \{$\wm,\ldz$\}, in such a way that marginalized uncertainties do not improve
if only one of the observables is measured.  Joint measurement can slightly improve
on this ``floor'' due to the difference in how the nuisance parameters affect the two.

\begin{figure}
\centerline{\epsfxsize=3.0truein\epsffile[80 520 320 700]{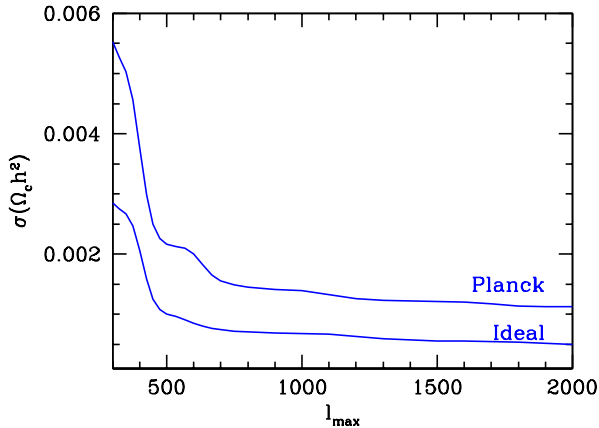}}
\caption{\footnotesize Uncertainty $\sigma(\wm)$ from CMB measurements with varying $\ellmax$,
for Planck sensitivity (Tab.~\ref{tab:planck}), and ideal measurement of unlensed CMB temperature
and polarization.}
\label{fig:lmax}
\end{figure}

One source of nuisance error arises from uncertainty in $\wm$ as measured from the primary CMB
or external data.
This is the dominant source of error throughout Tab.~\ref{tab:nuisance}, in
which we have assumed that the range of
multipoles from which cosmology can be extracted is limited to $\ellmax=2000$.  
This cut reflects an estimate of possible contamination from other secondaries and
foregrounds and is currently uncertain.
Fortunately
the dependence of $\sigma(\wm)$ on $\ellmax$ is not particularly sensitive strong
(see Fig.~\ref{fig:lmax}), provided that it exceeds the knee
at $\ellmax \sim 700$, corresponding to the trough between the second and third acoustic peaks.

Another conclusion from Fig.~\ref{fig:lmax} is that achieving cosmic variance limited
measurement of $E$-mode acoustic peaks at intermediate $l$ would significantly improve
$\wm$ constraints from Planck.
For example, an ideal measurement of \{$T,E$\} with $\ellmax\sim 500$ would
obtain $\wm$ constraints comparable to Planck, even though Planck will measure CMB temperature
anisotropy well into the damping tail.
We have found that the inability to extract $\wm$ constraints from the temperature damping tail
is due to confusion with the spectral index $n_s$; if $n_s$ is fixed rather than
marginalized, then Planck's $\wm$ constraints significantly improve.

As $\ellmax$ is increased in Fig.~\ref{fig:lmax}, the error on $\wm$ improves, but the error on
our second nuisance parameter $\ldz$ stays nearly constant.
This is because the degeneracy between $\ldz$ and $\tau$ in the unlensed CMB is broken only by
the reionization signal at low polarization multipoles.
Even at $\ellmax=2000$, the largest value we consider in this paper, the nuisance errors
are dominated by uncertainty in $\wm$.
This may suggest that $\ldz$ is never important as a nuisance parameter.   There are three
reasons why this may not be the case in practice.

First, errors on $\wm$ can be in principle be improved over CMB determinations
by external sources such as weak lensing
of galaxies  or even from the CMB itself through lens reconstruction, whereas  amplitude uncertainties
from reionization will likely remain. Second, we assume that large angle polarization foregrounds
can be perfectly removed to the cosmic variance limit through the frequency channels not used
for cosmology.  These foregrounds are now known to dominate the polarization signal in all bands
\cite{Speetal06}.

Third, our parameter forecasts so far have implicitly assumed sharp reionization, characterized by the
single parameter $\tau$, the total optical depth to recombination.
More general models of reionization can include additional parameters which degrade uncertainties
on $\tau$ \cite{KapOth02,HolHaiKapKno03}, and therefore our nuisance
parameter $\ldz$, beyond what is shown in Tab.~\ref{tab:nuisance}.

If we conservatively assume that the ionization history takes an arbitrary form,
uncertainties rise to 
$\sigma(\tau) = \sigma(\ldz) = 0.01$ \cite{HuHol03} and by Tab.~\ref{tab:theta}, the corresponding
nuisance errors are $\sigma_{\rm nuis}(\Theta_1) = \sigma_{\rm nuis}(\Theta_2) = 0.02$.
Comparing with Tab.~\ref{tab:nuisance}, this would be a comparable source of nuisance error
to $\wm$ for Planck, and would represent the dominant uncertainty for 
an ideal experiment that achieves cosmic variance on the $E$-modes
(c.f.~Fig.~\ref{fig:lmax}).

Finally the choice of $\ellmax$ also affects constraints on $\Theta_{1}$.  For example
scaling back to $\ellmax=1500$ would degrade $\sigma(\Theta_1)$ by $1.2$ for Planck and $1.6$
for the ideal measurement.

\subsection{Optimizing sensitivity to lensing $B$-modes}
\label{sec:optimization}

An important issue for upcoming polarization experiments is optimizing sky coverage
with the total integrated sensitivity fixed.  In general this is not a well-posed question since 
the sky coverage can be optimized with respect to systematic errors,
foreground contamination, sensitivity to the $E$-mode power spectrum,
sensitivity to tensor $B$-modes, detection of  lensing $B$-modes, $B$-mode reconstruction
of the lensing fields, or sensitivity
to cosmological parameters from the $B$-mode power spectrum.
Here, we consider only the last of these, with particular attention to how optimizing
the sensitivity is affected by non-Gaussian statistics.  

The integrated sensitivity $\mu$ has units of temperature
and is given by $\nu/\sqrt{NT}$ where $\nu$ is the instantaneous sensitivity
(in mK$\sqrt{\mbox{sec}}$) per Stokes parameter per detector,
$N$ is the number of detectors, and $T$ is the total integration time.  The 
noise variance per steradian in Eq.~(\ref{eq:NLdef}) is then given by 
$\Delta_P = \mu \sqrt{A}$ where $A$ is the survey area in steradians.

\begin{figure}
\centerline{\epsfxsize=3.0truein\epsffile[80 520 320 700]{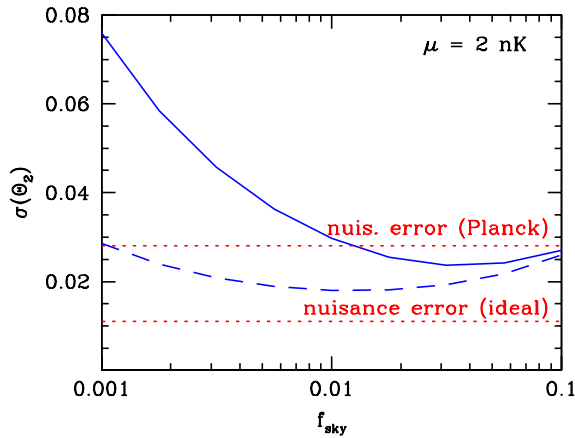}}
\caption{\footnotesize Dependence of the figure of merit $\sigma(\Theta_2)$ on $\fsky$, for fixed integrated sensitivity $\mu=2$ nK.
The solid line shows the result using non-Gaussian statistics for lensed $B$-modes (\S\ref{sec:pscov}); 
the dashed line shows the
result if Gaussian statistics are falsely assumed.  The horizontal lines represent the ``nuisance'' errors on $\Theta_2$
from imperfect measurement of \{$\wm$, $\ldz$\}, for Planck and for ideal measurement of the unlensed CMB 
(Tab.~\ref{tab:nuisance}).}
\label{fig:fsky}
\end{figure}

In \S\ref{sec:deobs}, we found that all cosmological constraints from the lensing $B$-mode power
spectrum are derived from the single observable $\Theta_2$.
Therefore, there is a natural figure of merit for optimizing sky coverage: the $1\sigma$
error $\sigma(\Theta_2)$.
In Fig.~\ref{fig:fsky}, we show $\sigma(\Theta_2)$ for varying $\fsky$, assuming fixed integrated
sensitivity $\mu=2$ nK.
Incorporating non-Gaussian statistics increases the optimal value of $\fsky$
by a factor of 3 relative to Gaussian, and significantly steepens the dependence of
$\sigma(\Theta_2)$ on  $\fsky$ in a manner which disfavors small $\fsky$.

We have shown the $\fsky$ optimization in detail for integrated sensitivity $\mu=2$ nK;
in general, the optimal $\fsky$ will scale with $\mu$ as $\fsky\propto\mu^{-2}$.
The same scaling is obtained assuming either Gaussian or non-Gaussian statistics; therefore,
the optimal patch size with non-Gaussian statistics incorporated is a factor of 3 larger than
the Gaussian value, independent of the integrated sensitivity.
Another way of stating the optimality criterion is: the optimal patch size is always chosen
so that the noise per steradian takes the value
\be
(\Delta_P)_{\rm optimal} = 4.7\mbox{ $\mu$K-arcmin}.
\label{eq:47}
\ee
This criterion makes no reference to the value of $\mu$ but does assume zero beam;
we have found that $(\Delta_P)_{\rm optimal}$ is nearly independent of beam size, provided
that $\theta_{\rm FWHM} \le 15$ arcmin.
(We note that if Gaussian statistics were falsely assumed for lensing $B$-modes, then one
would obtain $(\Delta_P)_{\rm optimal}=2.8$ $\mu$K-arcmin.)

\begin{figure}
\centerline{\epsfxsize=3.0truein\epsffile[80 520 320 700]{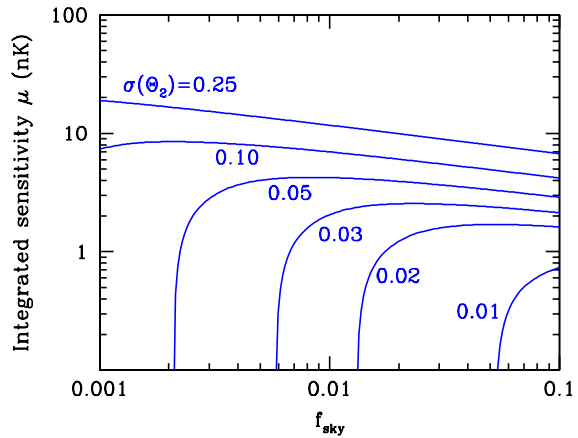}}
\caption{\footnotesize Level contours for the figure of merit $\sigma(\Theta_2)$ in the $\fsky$-$\mu$ plane,
assuming zero beam.}
\label{fig:optcontour}
\end{figure}

Another conclusion of the previous section was that, for sufficiently precise measurements of
lensing $B$-modes, the ability to extract cosmological information from the CMB alone
is primarily limited by
uncertainty in $\wm$ from the primary CMB and secondarily limited by reionization if the
ionization history is complex.
More precisely, when the measurement error $\sigma(\Theta_2)$ becomes as good as the the nuisance 
error $\sigma_{\rm nuis}(\Theta_2)$, then improved sensitivity to lensing $B$-modes serves mainly to
correlate low-redshift and nuisance parameters, rather than improving marginalized uncertainties on either.
In Fig.~\ref{fig:optcontour}, we have shown the dependence of $\sigma(\Theta_2)$ on total
sensitivity and sky coverage.

We now consider optimization of a deep ground-based polarization survey designed to complement Planck.
For the Planck prior on $\wm$, we found $\sigma_{\rm nuis}(\Theta_2) = 0.028$ (Tab.~\ref{tab:nuisance}).
Comparing with Fig.~\ref{fig:optcontour}, it is seen that a narrow-beam polarization survey with
integrated sensitivity $\mu$=1-2 nK, and covering a few percent of the sky, will achieve
$\sigma(\Theta_2) \approx \sigma_{\rm nuis}(\Theta_2)$ and is therefore nearly
optimal for extracting
cosmological information from lensing $B$-modes alone, within the limits of the Planck prior.

\begin{figure}
\centerline{\epsfxsize=3.0truein\epsffile[80 520 320 700]{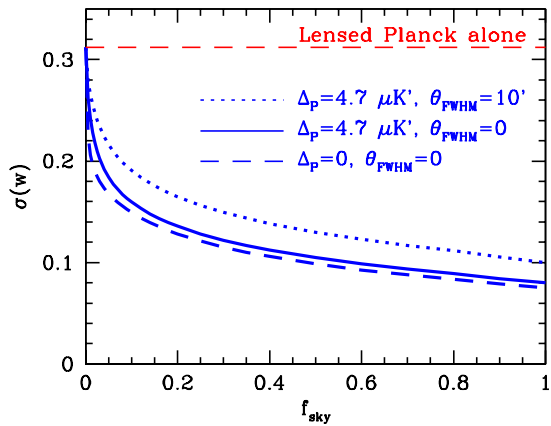}}
\caption{\footnotesize Uncertainty on the single low-redshift parameter $w$, marginalized
over high-redshift parameters, for Planck complemented by a deep survey with varying
sky coverage $\fsky$.  For the deep survey, we assume $\Delta_T = \Delta_P/\sqrt{2}$ and
consider three sensitivity levels as indicated.}
\label{fig:wfsky}
\end{figure}

\begin{table}
\begin{center}
\begin{tabular}{|c|c|c|c|c|}
\hline     Unmarg.         & Lensed Planck & Planck+Deep$_{5\%}$ &  Reference   &  Ideal  \\  \hline
$\sigma(w_0)$        &    0.31      &   0.18                 &  0.15   &  0.07   \\
$\sigma(w_a)$        &    0.65      &   0.38                 &  0.30   &  0.15   \\
$\sigma(\Omega_K)$   &    0.0076    &   0.0032               &  0.0025 &  0.0013 \\
$\sigma(\sum m_\nu)$ &    0.20     &   0.085                &  0.063  &  0.032  \\  \hline
\end{tabular}
\end{center}
\caption{\footnotesize Uncertainties on each of \{$w_0,w_a,\Omega_K,\sum m_\nu$\} separately with the others fixed,
and high-redshift parameters marginalized.
Here, ``Deep$_{5\%}$'' stands for a deep survey with $\fsky=0.05$, zero beam,
and $\Delta_P = \sqrt{2}\Delta_T = 4.7$ $\mu$K-arcmin (see Eq.~(\ref{eq:47})).
The reference survey is as in \S\ref{sec:exfisher} and covers $\fsky=0.10$, and ``Ideal'' refers to all-sky cosmic variance
limited \{$T,E,B$\} to $\ellmax=2000$. Forecasts here are from the direct Fisher approach.}
\label{tab:fact2}
\end{table}

This conclusion that a few percent of the sky is optimal for extracting the $B$-mode 
information alone is slightly modified once the $T$ and $E$ information from
Planck and the deep ground-based survey itself are considered.
The $T$ and $E$ measurements supply $\Theta_1$ and the nuisance parameters.   The first
modification is that if Planck succeeds in measuring $\Theta_1$ to the forecasted
$\sigma(\Theta_1)=0.05$ then it alone has lensing information that is comparable to
an $\fsky \sim 0.01$ $B$-mode survey.  The second modification is that as the deep survey
improves the nuisance parameters directly or through $\Theta_1$, the nuisance floor
on the $B$-modes correspondingly drops.  Both considerations favor slightly larger
$\fsky$.  To illustrate this we show in Fig.~\ref{fig:wfsky} the direct forecast on 
 a single low-redshift parameter, e.g.~$w$,
if Planck is complemented by a deep survey with varying $\fsky$.
For $\fsky$ less than the ``knee'' at $\fsky \sim 0.05$, the uncertainty in $w$ from the $B$-mode
measurement itself improves as $\fsky^{-1/2}$ but only becomes stronger than the lensing
constraint from Planck for $\fsky \gtrsim 0.01$.  For $\fsky \gtrsim 0.05$ improvements scale
more slowly as $\sim \fsky^{-1/3}$.  Half the total improvement comes from $\fsky < 0.1$.   
In fact, a deep survey with $\fsky = 0.05$ can improve Planck lensing uncertainties on any {\em one} of \{$w_0,w_a,\Omega_K,\sum m_\nu$\},
with the others fixed, by a factor of $\sim 2$ (Tab.~\ref{tab:fact2}) through measurement of the $B$-mode observable.   Moreover if lensing constraints from Planck prove impossible to extract
due to foreground and secondary contamination this improvement represents another 
factor of $\sim 2$ in errors.

\section{Discussion}
\label{sec:discussion}

We have provided a comprehensive study of the additional cosmological information 
supplied by lensed power spectra of the CMB temperature and polarization fields including
the non-Gaussian covariance between bandpower estimates.  This covariance originates from the
sample variance of the degree scale lenses on the CMB fields at smaller scales.  It is nearly
irrelevant for the temperature and $E$-polarization fields out to $\ellmax=2000$
due to the larger sample variance of
the unlensed CMB.   For the amplitude of the
$B$-polarization field, it increases the variance by up to a factor of $\sim 10$ and changes
the optimal observing strategy to one that covers a factor of $\sim 3$ times more sky area.

The impact of non-Gaussianity on parameter estimation 
as well as the net information content of the lensed spectra is  more subtle.  
These answers depend on the choice of parameters and the external priors associated
with them.
We have provided a framework of lensing observables that greatly simplifies these examinations.

In this framework, lensed CMB power spectra provide information on only two observables,
one which determines the lens power spectra at $l \sim 100$ 
associated with the \{$T,E$\}-fields  and one which determines it at $l \sim 500$ associated
with the $B$-field.  The observables are constructed from the  principal components of the lensing
power spectrum $C_l^{\phi\phi}$.  Non-Gaussianity is then automatically incorporated in the errors
on the observables which will eventually approach, but never exceed,  the sampling errors
of the lenses as the measurements improve.

This construction also illuminates the origin of parameter degeneracies which can rapidly
become the limiting source of uncertainties for parameters of interest.  Any combination of
parameters that leaves the lensing observables and the CMB at recombination fixed within
the errors cannot
be determined.
To illustrate these effects, we have isolated two parameters $\wm$ and $\ldz$ that determine
the shape and amplitude of the matter power spectrum respectively, and marginalized their
uncertainties assuming internal CMB determinations of each from the Planck satellite.
These become the limiting uncertainties once the observables are determined to the
several percent level and are only slowly improved as the lensing survey itself improves
the nuisance errors.
While $\wm$ constraints can be improved externally to the CMB, those on $\ldz$ are more 
difficult to improve and may be limited by our understanding of reionization.

There are also degeneracies within the space of the parameters of interest that control
the expansion rate and growth of structure at intermediate redshifts.
When taken one at a time, uncertainties on the parameters \{$w_0,w_a,\Omega_K,\sum m_\nu$\}
can be improved by a factor of $\sim 2-3$, relative to Planck alone, by a deep ground-based
polarization survey on 5-10\% of the sky.
However  \{$w_0, w_a$\} are nearly perfectly degenerate in the lensing observables
as are  \{$\Omega_K, \sum m_\nu$\} separately.  The degeneracy between two parameters
in each pair   is weakly broken
by the two observables.
For example, when errors on $\sum m_\nu$ are marginalized over $w_0$ they degrade
by a factor of 2 for the reference survey (see Fig.~\ref{fig:omnh2_w}).
However 
  sensitivity to the  \{$\Omega_K, \sum m_\nu$\} pair
  is much greater than to the dark energy parameters due 
 to the high redshift weights of the lensing observables.   
When combining lensed CMB power spectra with other more incisive 
probes of the dark energy, lensing
essentially fixes one well-defined combination of \{$\Omega_K,\wn$\} \cite{HuHutSmi06}.

Our conclusions have several caveats associated with them.  
The observables framework implicitly assumes that lensing is an independent
and additive source of cosmological information that may be combined with the
intrinsic CMB anisotropy.
An important exception to this statement occurs for tensor modes,
where lensing $B$-modes mask the intrinsic $B$-modes.
Forecasts for tensor modes should be made employing lensed power spectra
as a destructive contribution but here the Gaussian approximation suffices.
The conversion between instrumental noise and errors on the observables
depends only mildly on the fiducial model given current cosmological constraints
but we give a crude scaling in Appendix \ref{sec:fiducial}.

Secondly, we have considered only the information contained in the lensed power spectrum.
Beyond the power spectrum, non-Gaussianity from lensing allows a direct reconstruction of
the lensing fields \cite{HuOka01,HirSel02} which carries substantially more 
information that can break parameter degeneracies
\cite{Hu01c,KapKnoSon03}.  It may also allow ``de-lensing'' techniques
that recover the intrinsic $B$-modes from tensor modes 
\cite{KnoSon02,KesCooKam02,SelHir03}.
However techniques have yet to be developed that can remove systematics
and contamination at the levels required.

Thirdly, our parameter forecasts employ the Fisher matrix approximation.  It is well known
that Fisher matrix forecasts are not accurate along ill-constrained directions in the
parameter space.  Hence our results are only robust for quantities that lensed power
spectra constrain well.
Finally, we never consider CMB multipoles beyond $\ellmax=2000$ in this paper. Well
beyond this limit there is extra information on the high multipole structure of the lensing
field but this is likely to prove difficult to extract in the presence of other secondaries and foregrounds.

\section*{Acknowledgments}
We would like to thank Viviana Acquaviva, Carlo Baccigalupi, Lloyd Knox, Adrian Lee, 
Yong-Seon Song and Bruce Winstein for useful discussions.
We acknowledge use of the FFTW, LAPACK, and CAMB software packages.
KMS and WH were supported by the Kavli Institute for Cosmological Physics through the grant NSF PHY-0114422. WH was additionally supported by the grant 
DOE DE-FG02-90ER-40560 and the David and Lucile Packard Foundation.
MK was supported by the grant NSF PHY-0555689.

\appendix

\section{Goodness of Fit}
\label{sec:chisq}

\begin{figure}
\centerline{\epsfxsize=3.0truein\epsffile[60 360 320 700]{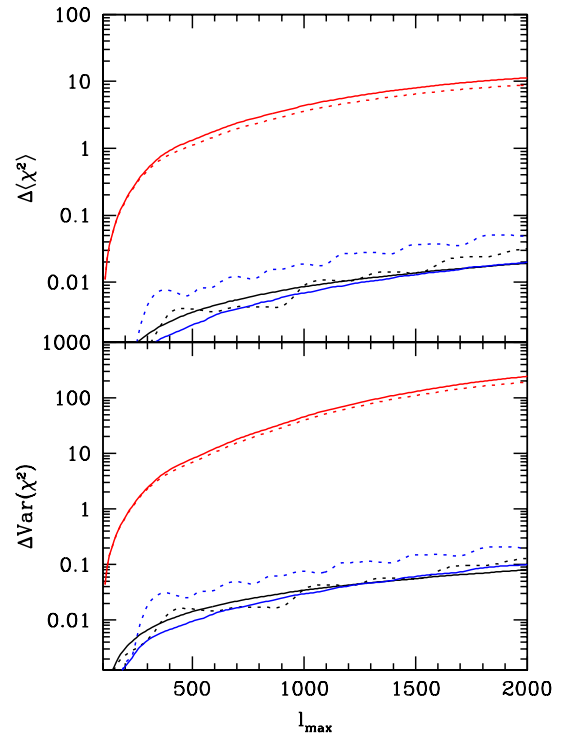}}
\caption{\footnotesize Non-Gaussian corrections $\Delta\langle\chi^2\rangle$ (top panel) and $\Delta\Var(\chi^2)$ (bottom panel),
as given by Eqs.~(\ref{eq:chisq1}),~(\ref{eq:chisq2}).
We assume sample variance limited measurements of $TT$ (black), $EE$ (blue), and $BB$ (red) up to maximum multipole $l_{\rm max}$.
For fixed $l_{\rm max}$, the corrections are nearly independent of the number of bands; we illustrate this by showing
a many-band fit (solid) and a one-band fit (dotted) for each power spectrum.
The corrections are negligible for $TT$ and $EE$ but important for $BB$, 
if $N_{\rm bands}$ is small and $l_{\rm max}$ is large.}
\label{fig:chisq}
\end{figure}

Given that lensing non-Gaussianity induces covariance in the band power estimates, it is interesting to 
ask  whether the correct cosmological model would be inferred
to be a bad fit to the observed bandpowers if non-Gaussian correlations were not included
in the $\chi^2$. This question was raised  in the context of the first-year WMAP analysis \cite{Speetal03}.  Correspondingly, we define the naive $\chi^2$ statistic as a sum over bands,
\be
\chi^2 = \sum_i \frac{(\hD_i - \langle\hD_i\rangle)^2}{\G_{ii}} \,,
\label{eq:chidef}
\ee
where the bandpower average $\langle\hD_i\rangle$ is computed using lensed power spectra, and
the Gaussian variance $\G_{ii}$ is computed assuming lensed power spectra and Gaussian statistics,
as in Eq.~(\ref{eq:Gdef}).  

The $\chi^2$ statistic defined by Eq.~(\ref{eq:chidef}) fully incorporates the effects of lensing at the power spectrum level
but neglects the non-Gaussian covariance between bandpowers.
With non-Gaussianity included, the distribution is no longer a perfect $\chi^2$, but acquires corrections
\bea
\langle\chi^2\rangle &=& N_{\rm dof} + \Delta\langle\chi^2\rangle \,,  \nn \\
\Var(\chi^2)         &=& 2N_{\rm dof} + \Delta\Var(\chi^2) \,,  \label{eq:chisq0}
\eea
where the excess contributions $\Delta\langle\chi^2\rangle$, $\Delta\Var(\chi^2)$ arise only from higher-point correlations
in the lensed CMB.  
In this appendix, we study the size of these contributions, as a way to quantify the impact of non-Gaussianity.

The first contribution in Eq.~(\ref{eq:chisq0}) can be written in terms of the bandpower covariance
defined in Eq.~(\ref{eq:GNdef}):
\be
\Delta\langle\chi^2\rangle = \sum_i \frac{\N_{ii}}{\G_{ii}}\,.
\label{eq:chisq1}
\ee
In contrast, the full non-Gaussian contribution to $\Var(\chi^2)$ is an eight-point correlation between CMB fields,
and the results of this paper do not permit every term to be computed.
However, if we make the approximation that the bandpowers $\hD_i$ are Gaussian variables, then it is given by:
\be
\Delta\Var(\chi^2) \approx 2 \sum_{ij} \frac{\N_{ij}\N_{ij} + 2\G_{ij}\N_{ij}}{\G_{ii}\G_{jj}}\,.
\label{eq:chisq2}
\ee
Since each bandpower is an average over many Fourier modes (see Eq. (\ref{eq:Ddef})), 
the central limit theorem implies that this should be an accurate approximation.
This general observation shows that, in the limit of wide bands, the bandpowers $\hD_i$ should always behave
as Gaussian variables; lensing simply induces a Gaussian covariance between the bandpowers.

We have found that the non-Gaussian contributions $\Delta\langle\chi^2\rangle$, $\Delta\Var(\chi^2)$ to
{\em unreduced} $\chi^2$ values are nearly independent of the number of bands or degrees of freedom.
In Figure \ref{fig:chisq}, we show these contributions for lensed $TT$, $EE$ and $BB$ 
power spectra, and for
varying $l_{\rm max}$, in two extreme cases: a ``many-band'' fit with $\Delta l = 10$, and a ``one-band'' fit
across all multipoles up to $l_{\rm max}$.
The non-Gaussian contributions are always negligible for $TT$ and $EE$; for $BB$ 
they are significant if the
number of bands is small and $l_{\rm max}$ is sufficiently large, but can be hidden if the fit is performed 
using many bands.
This is consistent with the discussion in \S\ref{sec:pscov}; non-Gaussianity is hidden when considering narrow $l$ 
bands, but appears as extra variance when estimated $BB$ power is averaged over a wide range in $l$.

In Figure \ref{fig:chisq}, we have computed $\Delta\langle\chi^2\rangle$ using Eq.~(\ref{eq:chisq1}), and
$\Delta\Var(\chi^2)$ using the approximation of Eq.~(\ref{eq:chisq2}).
To check this approximation, and the approximation that $\N_{ij}$ can be computed to lowest order
in $C_l^{\phi\phi}$, we have also computed $\langle\chi^2\rangle$, $\Var(\chi^2)$ using Monte
Carlo simulations of the lensed CMB, and find excellent agreement throughout Figure \ref{fig:chisq}.

\section{Fiducial Model Dependence}
\label{sec:fiducial}

Throughout this paper, we have presented results for the fiducial model of Eq.~(\ref{eq:fidmodel}) 
which has a low ionization optical depth and correspondingly a low $\sigma_8=0.73$.
For small deviations around the fiducial model, we have found that
the shape of the principal components (Fig.~\ref{fig:eigpp}) is unchanged, but the translation between
the noise level and uncertainties $\sigma(\Theta_i)$ in the lensing observables 
(Fig.~\ref{fig:sigmatheta}) can be affected.  
Denoting the uncertainty at noise level $\Delta_P$ by $\sigma(\Theta_i;\Delta_P)$,
we find the following rough scaling, expected from signal-to-noise considerations
assuming that the unlensed CMB is fixed:
\bea
\sigma(\Theta_1;\Delta_P) &\approx& \left( \frac{C_{l_{K1}}^{\phi\phi}}{C_{l_{K1},{\rm fid}}^{\phi\phi}} \right)
                                           \sigma_{\rm fid}(\Theta_1;\Delta_P)\,,  \\
\sigma(\Theta_2;\Delta_P) &\approx& \sigma_{\rm fid} \left[\Theta_2; 
         \left( \frac{C_{l_{K2}}^{\phi\phi}}{C_{l_{K2},{\rm fid}}^{\phi\phi}} \right)^{-1/2} 
         \!\!\!\!\!\!\!\! \Delta_P \right] \,.  \nn
\eea
Here, $l_{K1}=114$, $l_{K2}=440$ are the median multipoles from \S\ref{sec:pcomp}.
The scaling for $\Theta_1$ follows from considering the unlensed CMB as a fixed noise source 
whereas for $\Theta_2$ it follows from direct signal-to-noise scaling.

The optimal noise level from Eq.~(\ref{eq:47}) for measuring lensing $B$-modes scales roughly as
$(C_{l_{K2}}^{\phi\phi}/C_{l_{K2},{\rm fid}}^{\phi\phi})^{1/2}$ for the same reason.

\vfill
\bibliography{lenscov}

\end{document}